\documentclass[preprint]{aastex631}

\usepackage{inputenc}
\usepackage{amsmath}

\newcommand{\be}{\begin{equation}}
\newcommand{\ee}{\end{equation}}
\newcommand{\rsun}{\mathrm{R}_\odot}

\shorttitle{Coronal density structuring from Comet Lovejoy}
\shortauthors{Nistic\`o et al.}

\begin{document}

\title{Probing the Density Fine Structuring of the Solar Corona with Comet Lovejoy}

\author[0000-0002-0786-7307]{Giuseppe Nistic\`o} 
\author[0000-0002-9207-2647]{Gaetano Zimbardo}
\author[0000-0002-8399-3268]{Silvia Perri}
\affiliation{Dipartimento di Fisica, Universit\`a della Calabria, Via P. Bucci, Cubo 31 C, Arcavacata di Rende, Cosenza, 87036, Italy}
\affiliation{National Institute for Astrophysics, Scientific Directorate, Viale del Parco Mellini 84, I-00136, Roma, Italy}

\author[0000-0001-6423-8286]{Valery M. Nakariakov}
\affiliation{ Centre for Fusion, Space and Astrophysics, University of Warwick, Coventry, CV4 7AL, UK}
\affiliation{Centro de Investigacion en Astronomía, Universidad Bernardo O’Higgins, Avenida Viel 1497, Santiago, Chile}

\author[0000-0003-3306-4978]{Timothy J. Duckenfield}
\affiliation{Centre for Mathematical Plasma Astrophysics (CmPA), KU Leuven, Celestijnenlaan 200B bus 2400, 3001 Leuven, Belgium}

\author[0000-0001-7312-2410]{Miloslav Druckm\"uller}
\affiliation{Faculty of Mechanical Engineering, Brno University of Technology, Technick\'a 2, 616 69 Brno, Czech Republic}

\begin{abstract}
The passage of sungrazing comets in the solar corona can be a powerful tool to probe the local plasma properties. 
Here, we carry out a study of the striae pattern appearing in the tail of sungrazing Comet Lovejoy, as observed by the Atmospheric Imaging Assembly (AIA) aboard the Solar Dynamics Observatory (SDO) during the inbound and outbound phases of the comet orbit. We consider the images in EUV in the 171 {\AA} bandpass, where emission from oxygen ions O$^{4+}$ and O$^{5+}$ is found.
The striae are described as due to a beam of ions injected along the local magnetic field, with the initial beam velocity decaying because of collisions. Also, ion collisional diffusion contributes to ion propagation. Both the collision time for velocity decay and the diffusion coefficient for spatial spreading depend on the ambient plasma density. A probabilistic description of the ion beam density along the magnetic field is developed, where the beam position is given by the velocity decay and the spreading of diffusing ions is described by a Gaussian probability distribution. Profiles of emission intensity along the magnetic field are computed and compared with the profiles along the striae observed by AIA, showing a good agreement for most considered striae. The inferred coronal densities are then compared with a hydrostatic model of the solar corona. The results confirm that the coronal density is strongly spatially structured.
\end{abstract}


\section{Introduction} \label{sec:intro}

Daily observations of the white-light solar corona with the Large Angle and Coronal Spectrometer (LASCO) aboard the Solar and Heliospheric Observatory (SoHO) allowed to discover more than three thousands comets plunging into the harsh solar atmosphere \citep{Battams2017}. Because of the proximity of their perihelion, such comets are usually referred to as sungrazers. \citet{Jones2018} gave a more tight classification of comets based on their perihelion distance ($q$) and distinguished between near-sun comets ($q<66.1~\rsun$ from the centre of the Sun), sunskirting ($3.45~\rsun <q<33.1~\rsun$), sungrazing ($q=1.0-3.45~\rsun$), and sundiving ($q<1.0~\rsun$) comets. Based on their orbits, sungrazing comets are grouped into families since they are fragments of a common progenitor, like, the Kreutz group. Commonly, sungrazing comets appear in coronagraphs as small and fast-moving bright dots with a short tail, but some of them are outstanding events, with very long tails and bright comas. When approaching the Sun, the majority of these comets are completely dissolved within a couple of solar radii from the Sun's surface. Nevertheless, two comets were observed in the FoV of the Atmospheric Imaging Assembly (AIA) of the Solar Dynamics Observatory (SDO), hence at distances below 0.7 $\rsun$ from the solar surface in the extreme ultra-violet wavelengths (EUV): comets C/2011 N3 \citep{Schrijver2012} and C/2011 W3 (Lovejoy) \citep{Downs13}. Another comet approaching the Sun in November 2013, C/2012 S1 (ISON), disregarded expectations and failed to show EUV signatures of its passage in the corona, probably because of the lower size of its nucleus that could not withstand the intense heat from the Sun \citep{Bryans2016}.

The transit of Comet Lovejoy in December 2011 was exceptional. The comet crossed the corona from East to  West, with the perihelion located behind the Sun as observed from Earth, at only $\sim0.2~\rsun$ from the Sun's surface. The images recorded by SDO in the different EUV channels of AIA showed a trail of striations, also referred to as {\it striae}. \citet{Bryans2012} associated the EUV striae to the temporal evolution of the excited states of oxygen ions released by the comet nucleus and guided along the local magnetic field lines, which were emitting in EUV as a consequence of the increase in their relative abundance. \citet{Downs13} used sophisticated MHD simulations to reproduce the observations and provided estimates of the magnetic field and coronal density of the corona.   
\citet{McCauley2013} focused on the post-perihelion observations of Lovejoy from AIA and included in their analysis X-Ray Telescope (XRT) data from Hinode. They determined the ionisation stages in the observed striae and determined other physical parameters, such as the out-gassing rate of the nucleus and the mass of the comet. \citet{Raymond2014} also analysed the striae left by the comet during the outbound phase, estimated the speed of propagation and broadening of the striae along the local magnetic field, and used an MHD model and the theory of pick-up ions to compare the measurements with theoretical expectations.
They found that cometary ions outline flux tubes with diameters of about 4000 km and that density variations between neighbouring flux tubes reach at least a factor of six on a scale of few thousand kilometers. The measurements were compared with an MHD model and the theory of pick up ions to show that part of the energy of the cometary ions as they isotropize  goes to Alfv\'en waves. However, \citet{Raymond2014} did not consider the Coulomb interaction between oxygen ions and the background coronal plasma when estimating velocities and densities.

In general, the relevance of comets relies on the possibility to exploit them as natural probes of the inner heliosphere and solar corona. In fact, even before the advent of exploration by spacecraft, the direction of plasma tails of comets observed from Earth during their transit was thought as an evidence of solar wind flow \citep{1951ZA.....29..274B,1958ApJ...128..664P}. Several studies on comet observations aimed at determining the physical condition of the interplanetary medium (solar wind speed, density, temperature etc.) and assessing the physical processes at work in the coupling between the solar wind and comet tails \citep{2007ApJ...668L..79V,2015ApJ...812..108D,2018A&A...615A.143N}. Along the same lines, sungrazing comets have been used to infer the conditions of the solar corona but at distances much further from the photosphere, using the SoHO/LASCO and UVCS instruments \citep{2007P&SS...55.1021B, 2018ApJ...858...19R}. In the case of Comet Lovejoy, 
the importance in studying the EUV striae concerns the possibility to diagnose the physical conditions of the low solar corona at distances inaccessible to spacecraft. Even Parker Solar Probe, which will attain a minimum distance of 9.8 $\rsun$, will also not be able to get as close to the solar surface as Comet Lovejoy did.

In this work, we aim at studying the time evolution of the striae by modeling the propagation of oxygen ions along a magnetic flux tube. The length of a stria is determined by the ion beam velocity, which decays because of collisions, and the ion collisional diffusion. By determining the parameters characterizing the beam slowing down and the collisional dispersion, we can get estimates of the ambient plasma density at the location where the striae are formed.
The paper is organised as follows: in Section \ref{sec:obs} we present the observations of Lovejoy from SDO/AIA; in Section \ref{sec:modeling} we show how the striae of Comet Lovejoy are modeled; Section \ref{sec:analysis} discusses the analysis of the SDO/AIA data and the comparison with the numerical modeling. Finally, the discussion and conclusions are given in Section \ref{sec:disc}.

\section{Observations} \label{sec:obs}

\begin{figure}[htpb]
    \centering
        \includegraphics[width=1.0\textwidth]{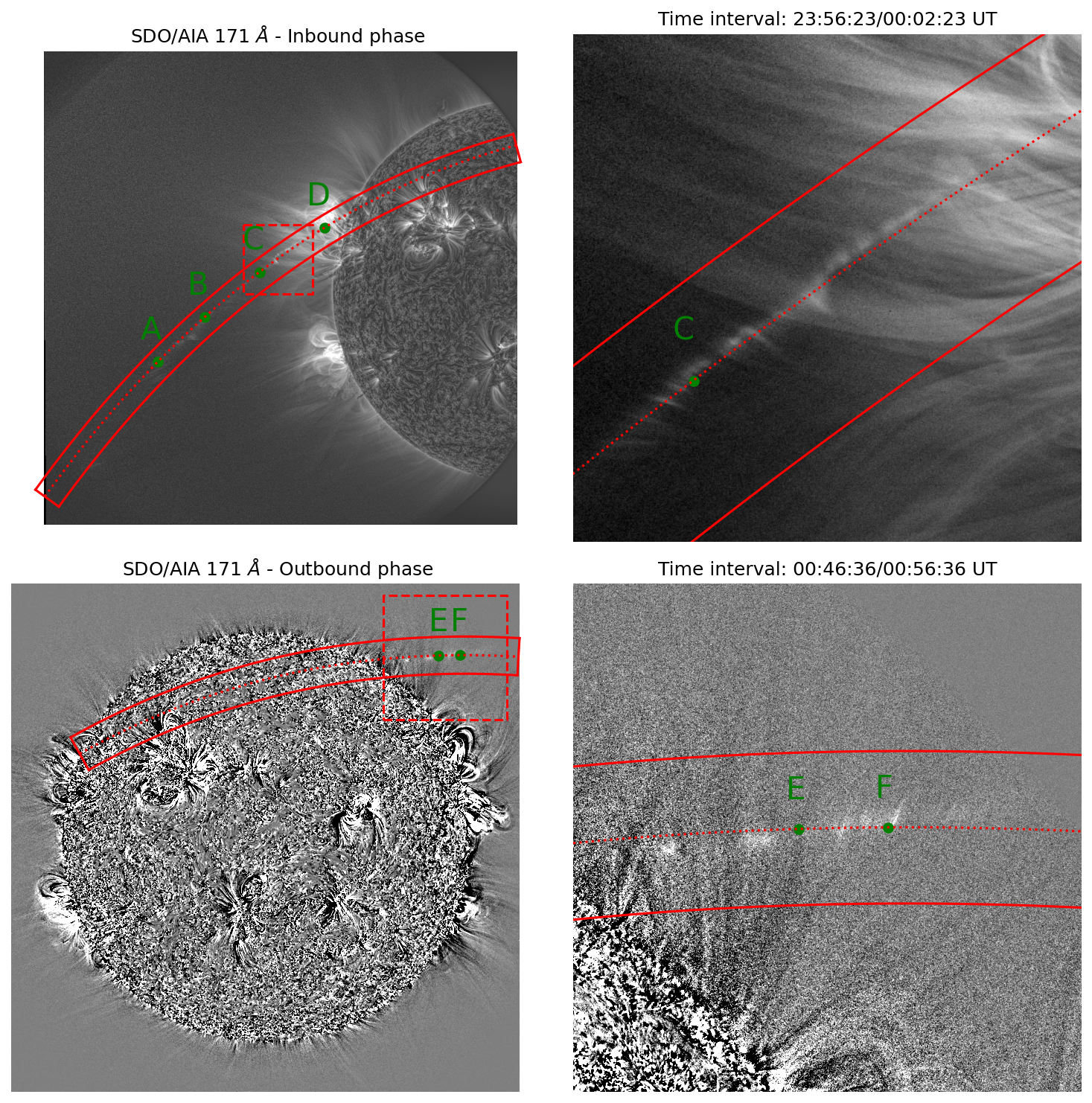}
    \caption{Left panels: Composite images taken at different times with SDO/AIA in the 171 channel and showing the tail of striae left by Comet Lovejoy enclosed within two continuous red lines before (top panels) and after (bottom) the transit at the perihelion. The orbit of the comet is shown as a dotted red line. Right panels: blow-up view of the regions where striae are formed. The corresponding regions are outlined with a dashed red box in the left panels. The filled green circles labelled by letters mark the locations where intensity profiles are extracted and plotted in Fig. \ref{fig:int_vs_cpos}.}
    \label{fig:context_image}
\end{figure}

\begin{figure}
    \centering
        \includegraphics[width=1.0\textwidth]{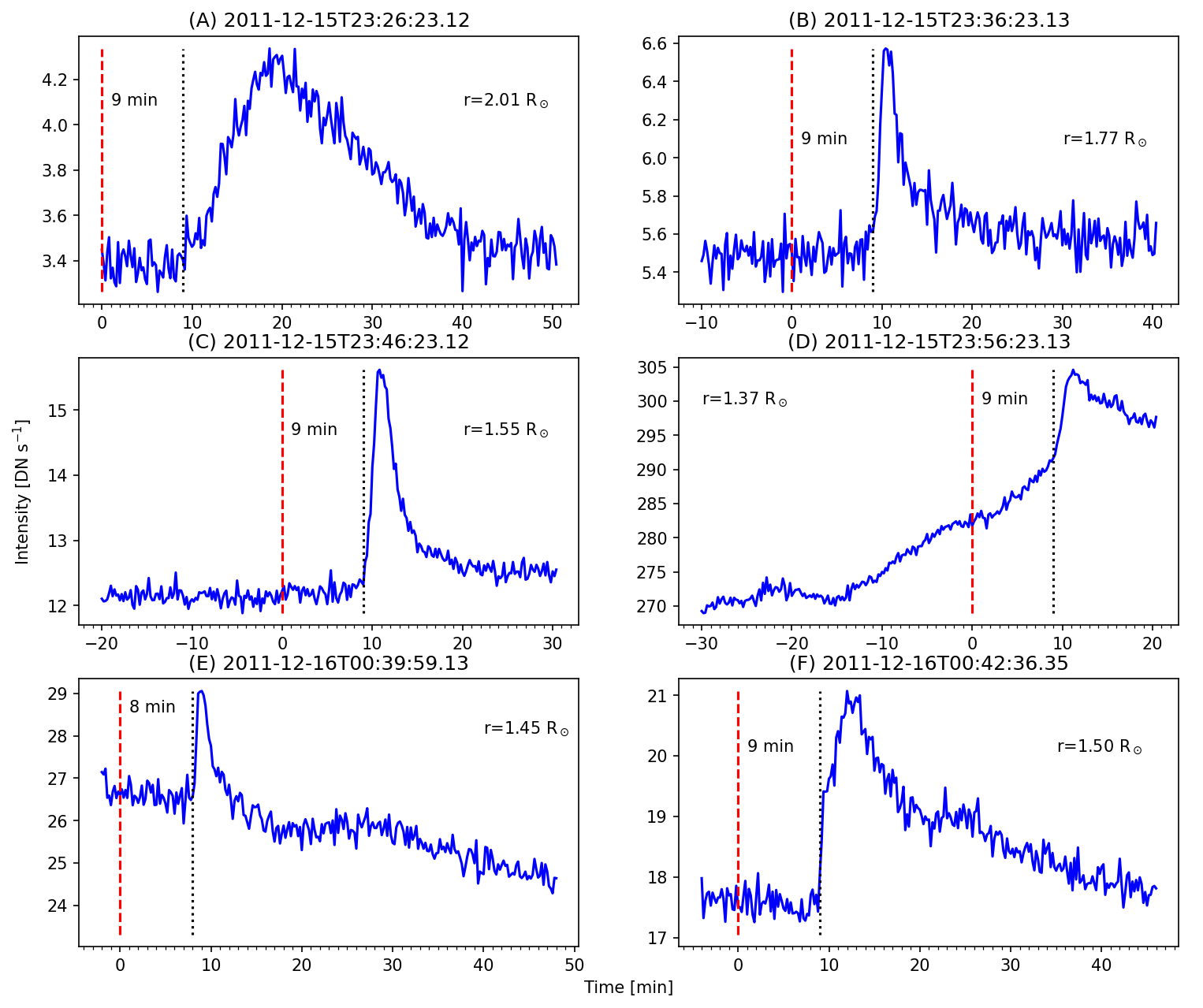} 
    \caption{Intensity profiles vs. time taken at six different locations along the path of the comet. The dashed red line marks the time instant of the transit of the comet nucleus. The vertical dotted line indicates the moment at which the EUV emission starts to increase, hence when the oxygen ions starts to emit. The time lag is calculated between these two vertical lines and must be corrected for the light travel time to the observer.}
    \label{fig:int_vs_cpos}
\end{figure}

The perihelion transit of Comet Lovejoy was seen by different space observatories. SDO/AIA recorded the approach to the perihelion between 23:25 and 00:16 UT on December 15--16, 2011. For that occasion, the pointing axis of SDO/AIA was offset by about half degree to observe the comet entering in the corona. Afterward, the comet was seen to re-emerge from the perihelion on the West side between 00:40 and about 01:00 UT.
Fig. \ref{fig:context_image} shows a composite image taken in the 171 \AA~ channel of SDO/AIA showing the inbound (top panels) and outbound phases of the comet (bottom panels). The images in the top panels are processed with the Noise Adaptive Fuzzy Equalization tron (NAFE) algorithm \citep{2013ApJS..207...25D}.
The trajectory of the comet is superimposed onto the images as a dashed red line and is computed with SPICE: the SPICE kernel file was downloaded from the Horizons System \footnote{\url{https://ssd.jpl.nasa.gov/horizons/app.html\#/}} and read with the routines of the Spiceypy Python package \citep{Annex2020}. FITS file images were download from the JSOC service and read using the \texttt{read\_sdo.pro} routine in SolarSoftWare (SSW)/IDL.
We did not process the FITS files with the standard routine \texttt{aia\_prep.pro} because the images of the sequence were not correctly calibrated based on the FITS header information. The images were taken with a 12-second cadence (which was almost uniform during the observations) and a pixel size $\Delta_{\mathrm{pix}}\approx$0.56$"$. Some persistent jitter in the inbound phase data was removed by co-aligning each image of the sequence to the middle one taken at 23:51:36.58 UT.
As shown in Fig. \ref{fig:context_image}, the passage of the comet inside the solar corona left a trail of EUV striae. The striae appeared a minute after the transit of the comet nucleus. In Fig. \ref{fig:int_vs_cpos} we show the intensity profiles versus time extracted from six different locations along the projected path followed by the comet. In each panel, the vertical dashed red line marks the time instant when the comet nucleus passed through the point, the dotted black line indicates the time when the EUV intensity starts to increase. Indeed, the stria signature is found as an increase, more or less sharp, in the EUV intensity. The time-lag varies between 8--9 minutes. These time-lags have to be corrected for the light travel time from the Sun to the Earth, which is about 8 min. After correction, the definitive time-lag between the passage of the comet and the appearance of an EUV stria is within at most one minute, thus confirming the estimate provided in \citet{McCauley2013} in different EUV bands. 
The striae appear different in length and brightness, reflecting the amount of emitting oxygen ions, which in turn depend on the local density of the corona. In the next Section we explain how the striae of Comet Lovejoy are modeled, in order to calculate intensity profiles and compare them with those obtained from the observations.

\section{Modeling of EUV Striae}\label{sec:modeling}

\subsection{Theoretical model}

\begin{figure}[htpb]
    \centering
    \includegraphics[width=.8\textwidth]{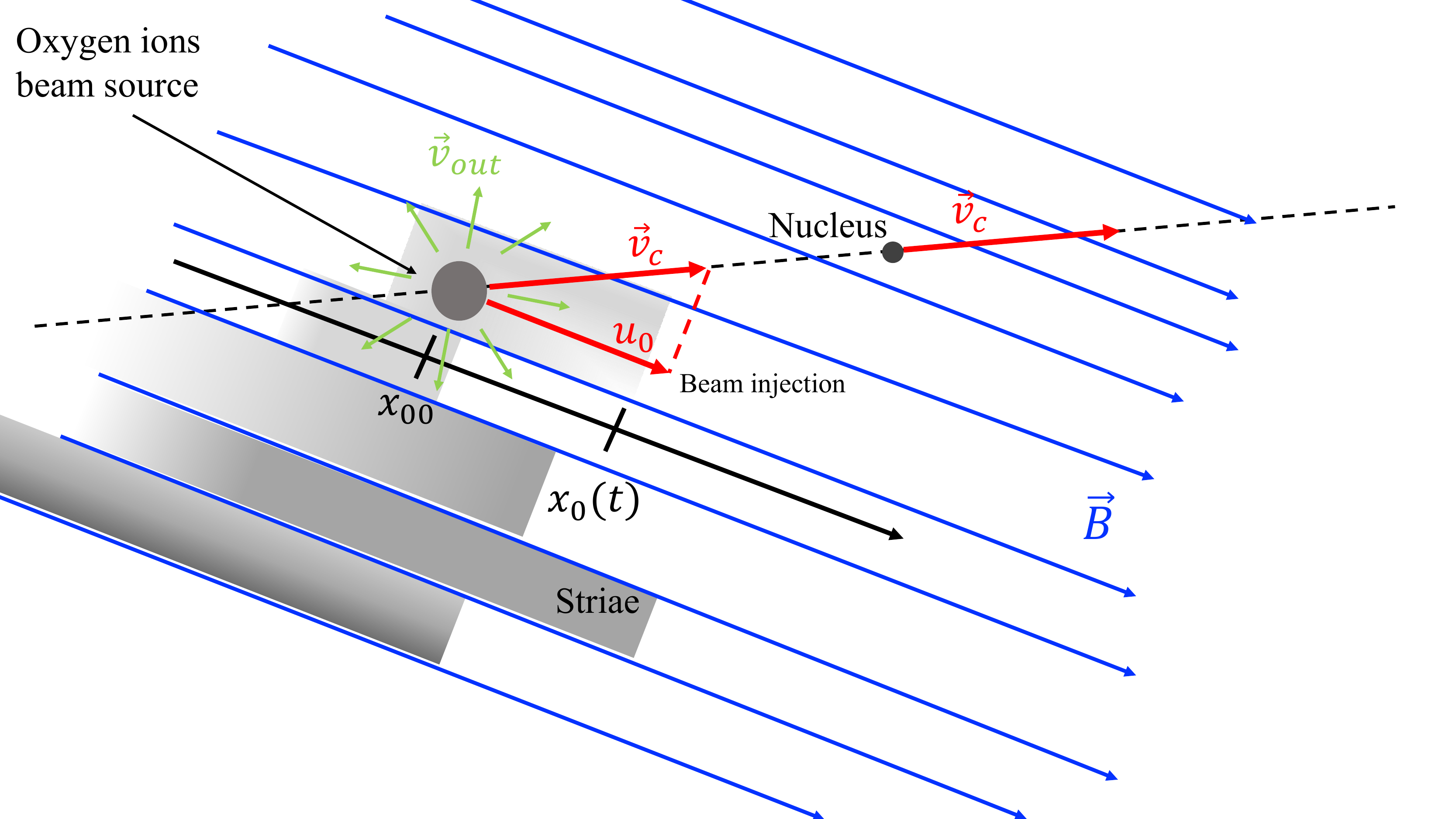}
    \caption{Scheme of the mechanism for the formation of the striae}
    \label{fig:fig1}
\end{figure}

Following \citet{Downs13}, we assume that the cometary material is extracted from the comet during its travel through the solar corona, and that this material is rapidly ionized.
Once oxygen and other atoms are ionized, they are captured by the magnetic field and start to spiral around the straight field lines, moving along them with a velocity $u_0$ that can be estimated as the projection of the ion velocity along the magnetic field, $u_0 = {\bf v}_i\cdot{\bf B}/B$. A schematic representation is reported in Fig. \ref{fig:fig1}. The initial ion velocity can be seen as the sum of the comet velocity ${\bf v}_c$, and the ion outflow velocity ${\bf v}_{out}$, which is usually modeled as radial with respect to the comet nucleus \citep{Bryans2012}: ${\bf v}_i = {\bf v}_c + {\bf v}_{out}$. 

As time goes on, oxygen ions reach higher ionization levels until they emit in the 171 \AA~line and are detected by SDO/AIA. This happens mainly for the ionization states O$^{4+}$ and O$^{5+}$ \citep{Downs13,Pesnell14} and implies that oxygen ions will be visible in 171 \AA~only some time after the comet passage, as discussed in the previous section. Later, these ions will be further ionized to O$^{6+}$ and more (e.g., by photioniosation, charge exchange, electron and proton collisions), so that emission in the 171 \AA~line fades out. In the meanwhile, oxygen ions move along the magnetic field and slow down because of collisions. They are also undergoing a diffusive spreading because of their disordered motion, due to the outflow velocity ${\bf v}_{out}$ and to collisions \citep[][Supplementary Material]{Downs13}.

The cloud of oxygen ions is spreading in the corona in three dimensions, however, since the Larmor radius is only a few km, which is much smaller than the striae length, we limit ourselves to a one dimensional modeling with the $x$ direction along the magnetic field, also  considering that (i) the spreading along the line-of-sight can be taken into account by integrating along the line of sight, something which is implicit in the EUV observations; and (ii) the spreading in the plane-of-sky and perpendicular to the $x$ direction is taken into account by actually summing the EUV signal over 4--5 pixels in that direction (see Section \ref{sec:analysis}), which also allows to improve the signal-to-noise ratio. 

Therefore, we propose a simple modeling of the beam of oxygen ions, which are treated as test-particles.  The initial velocity along the mean field direction decreases because of collisions as
\be
u(t) = u_0 e^{-t/\tau_{\rm slow}} \, ,
\ee
where $\tau_{\rm slow}$ is the time for the collisional slowing down of a suprathermal ion beam in a field plasma (see below). 
We consider that ${\bf v}_{out}$ is symmetric, $<{\bf v}_{out}>=0$, and that for the beam $u_0$ is given by the average of ${\bf v}_i$, so that $u_0 = {\bf v}_c \cdot {\bf B}/B$. Then, the position of the beam head $x_0(t)$ is
\be
x_0(t) = x_{00} + u_0 \tau_{\rm slow}\left[1-\exp(-t/\tau_{\rm slow})\right] 
\label{eq_beam}
\ee
where $x_{00}$ is the starting position of the beam, just after the release of ionized material from the comet into the solar corona. The above expression shows that, basically, the length of a stria is given by $u_0 \tau_{\rm slow}$, to which the spreading due to collisional diffusion has to be added. Here, we study the distribution of material along the magnetic field by a probabilistic description taking into account both the beam \lq\lq ordered\rq\rq motion and the ion diffusive spreading. By using a methodology adopted for the transport of energetic particles accelerated at shock waves \citep{Ragot97,Perri08,Zim18}, we model the number density $n(x,t)$ of ions along the magnetic field as
\begin{equation}
        n(x,t) =  \int_0^t dt' \int_{-\infty}^{+\infty} dx' Q(x',t',t) P(x-x', t-t')
\label{eq_density}
\end{equation}
where the source of oxygen ions is given by
\begin{equation}
        Q(x',t',t) =   \eta(t)\Phi(t') \delta(x'- x_0(t')) \, ,
\end{equation}
and $P(x-x', t-t')$ is the probability of observing a particle in $(x,t)$ if it was emitted in $(x',t')$. 
Here, $Q(x',t',t)$ represents the number of ions emitted in the position $x'$ at time $t'$. The motion of the beam is given by $x_0(t')$, i.e., by Eq. \eqref{eq_beam}, and the delta function implies that the source is located at $x'=x_0(t')$. Further, $\Phi(t')$ models the ion injection from the comet; if injection were sharp and punctual, this could be modeled as $\delta(t'-t_0)$ \citep[see, e.g.,][]{leRoux19}. However, here we have to take into account that the emission of oxygen ions from the comet is a multi-step process that includes solid particles and dust leaving the comet; molecule sublimation and ionization; and then oxygen ions reaching the ionization states O$^{4+}$ and O$^{5+}$. These steps require a time of several tens of seconds, as evidenced by the time lags shown in Figure \ref{fig:int_vs_cpos}, also depending on the coronal density and on the size of the dust grains, and cannot be well described by a sharp injection. Instead, we model the ion injection as a Gaussian function delayed in time by $t_0$:
\be
\Phi(t') = \frac{\Phi_0}{\sqrt{2\pi} {\sigma_{t_0}}} \exp\left[ -\frac{(t'-t_0)^2}{2\sigma_{t_0}^2}\right] \, .
\ee
In other words, $t_0$ represents the average time for oxygen ions to reach the O$^{4+}$ and O$^{5+}$ states, after the comet transit, and $\sigma_{t_0}$ the typical half-duration of the injection process. It would be possible to consider an injection process which is not point-like but also spreads out in space. We reserve this for future work.

In addition, we introduce in $Q(x',t',t)$ a factor $\eta(t)$ which is not strictly related to ion injection, but rather is a decay factor which corresponds to the ion beam ageing. Here, $\eta(t)$ describes the fact that after some time the oxygen ions pass from O$^{5+}$ to O$^{6+}$ and therefore the emission in the line {171 \AA} becomes negligible. Keeping a certain analogy with energetic particle transport, this effect is often modeled as a loss term in the transport equation, i.e., $\partial f/\partial t \simeq -f/\tau_{L} $ with $f$ the particle distribution function and with $\tau_{L}$ the loss time \citep[e.g.,][]{Perri16,leRoux19}. Such a loss term leads to an exponential decay. Therefore, we model this factor as $\eta(t)= \exp(-t/\tau_{L})$, with $\tau_{L}$ the ion lifetime which is obtained from \citet{Pesnell14}, see below. 

The probability $P(x-x', t-t')$ is given by the Gaussian propagator, which is appropriate in the case of collisional diffusion:
\begin{equation}
        P(x-x', t-t') =  \frac{1}{\sqrt{4\pi D_{xx}(t-t')}} \exp{\left[ -\frac{(x-x')^2}{4 D_{xx} (t-t')} \right]}
\label{eq_propag}        
\end{equation}
where the diffusion coefficient along the $x$ direction is given by $D_{xx} = \frac{1}{3}v_{\rm ran}^2\tau_D$. Here, $v_{\rm ran}$ is the disordered ion velocity after the ion \lq\lq first\rq\rq collision, and $\tau_D$ the dispersion collision time (see discussion below). 
We note that the oxygen ion beam represents a strongly non-thermal distribution which can be unstable, so that wave-particle interaction can also contribute to the ion slowing down and diffusive spreading \citep[e.g.,][]{Raymond2014}. Here, however, we overlook this possibility and concentrate on collisional effects.
By integrating Eq. \eqref{eq_density} over $x'$ and exploiting the delta function, we obtain
\begin{equation}
        n(x,t) =  \eta(t) \int_0^t dt' \Phi(t') \frac{1}{\sqrt{4\pi D_{xx}(t-t')}} \exp\left\{-\frac{\left[ x - u_0\tau_{\rm slow}(1-\exp(-t'/\tau_{\rm slow}))\right]^2}{4D_{xx}(t-t')}\right\}.
        \label{eq:eq_density_final}
\end{equation}
Further numerical integration over $t'$ gives the density profiles which can be used for comparison with the striae observations. 

\subsection{Collision times for oxygen ions and model parameters}

The above modeling of the oxygen O$^{4+}$ and O$^{5+}$ density along the striae depends on a number of parameters. Keeping in mind that several simplifications are needed in order to obtain a manageable treatment, these parameters are determined as follows.

The initial beam speed along the magnetic field $u_0$ is obtained from the projection of the comet speed on the magnetic field direction; because of projection effects, the angle $\theta$ between $\bf B$ and ${\bf v}_c$ is not well known. We note that \citet{Downs13} consider that $u_0$ may be in the range 100--200 km s$^{-1}$, while \citet{Raymond2014} estimate a velocity parallel to the magnetic field of 183 km s$^{-1}$ for some striae during the egress of comet Lovejoy. Here, we consider similar values of $u_0$, of the order of $u_0=100$ km s$^{-1}$, which are determined also with the help of the observed motion of the stria front. 
We consider that the ion outflow velocity, assumed to spring radially out of the comet, can also cause a significant spread in the value of $u_0$.

The temporal evolution of various ion fractional abundances has been studied by \citet{Pesnell14}; for O$^{4+}$ and O$^{5+}$, which give rise to the emission lines O$_{ V}$ and O$_{ VI}$, they find for an electron density $n_e = 10^7$ cm$^{-3}$ a rise time of about 50 and 200 s, respectively for O$^{4+}$ and O$^{5+}$, and an exponential decay time of 300 and 800 s, respectively. Since both ion species contribute to the {171 \AA}  emission line, considering that each oxygen ion goes through successive ionization states,  we assume a lifetime for UV emission of $\tau_L \simeq 10^3$ seconds, i.e., roughly the sum of the decay times for O$^{4+}$ and O$^{5+}$. On the other hand, the rise times reported by \citet{Pesnell14} and the time lags inferred from Figure \ref{fig:int_vs_cpos} are used to model the temporal evolution of emitting ion injection, that is, the delay time $t_0$ after the comet transit and the typical half-duration of the injection process: we assume initial values of $t_0 = 60$ s and $\sigma_{t_0} = 80$ s, and then adjust these values to best match each stria profile (see Table \ref{tab:tab2}).

For the elongation and spreading of the oxygen ion beam, basic parameters are the collision times $\tau_{\rm slow}$  and $\tau_D$, see Eqs. \eqref{eq_beam} and \eqref{eq_propag}. 
Because of the nature of Coulomb collisions, which implies a cross-section decreasing as $U^{-4}$, with $U$ the relative speed between colliding particles \citep[e.g.,][]{2005ipp..book.....G}, the time $\tau_{\rm slow}$  for the slowing down of a suprathermal ion beam and the time $\tau_D$ for dispersion of suprathermal particles are longer than the collision time of thermal particles  \citep[e.g.,][]{Krall,Downs13}.  
In turn, the suprathermal ions thermalize with the ambient plasma on a longer time scale than $\tau_{\rm slow}$ and $\tau_D$ \citep{Krall}. In this case, for an ion beam whose velocity is larger than the ion thermal velocity but slower than the electron thermal velocity, the following expression can be used \citep{Krall,Downs13}
\be
\tau_{\rm slow} \approx \frac{{m_T}^2}{4\pi n_e e^2 {q_T}^2 \left[ \displaystyle{ \left(1+\frac{m_T}{m_p}\right)\frac{1}{U^3}+\frac{4}{3\sqrt{\pi}}\left(1+\frac{m_T}{m_e}\right) \left(\frac{m_e}{2\kappa T_e} \right)^{3/2} } \right]\ln\Lambda} \, ,
\label{eq:tau_slow}
\ee 
 with $m_T$ and $q_T$ the mass and charge of the suprathermal beam particles, $U$ the speed of suprathermal  particles, $e$  the elementary charge, $m_p$ and $m_e$ the masses of proton and electron, $\kappa$ the Boltzmann's constant, $\ln\Lambda \simeq 22$ the Coulomb's logarithm. We have also assumed a coronal electron temperature $T_e \sim 1.5$ MK. An initial ion speed of the order of $U\simeq v_c\simeq 500$ km s$^{-1}$ could be taken. Actually, when computing the density of each stria we considered the comet speed $v_c$ when it created the stria at a certain distance from the Sun $r_c$. We note that $U$ is assumed to correspond to the comet's speed because this is the initial ion speed which is involved in the collision processes, even if the average beam speed along the magnetic field $u_0$ is lower. In other words, we assume that gyromotion is not modifying the collision time since the gyroradius is very much larger than the inter-particle distance. 
This collision time depends on the electron density $n_e$, which is kept as a free parameter to be determined by the fit of the observed emission profiles.

Now, we consider that after the first collision the ions of the beam move in random direction, that is, are subject to a diffusive motion. In such a scenario, we can assume that 
the random velocity of dispersive motion $v_{\rm ran}$ decays exponentially from the initial speed of the ions $\sim v_c$ to the thermal speed of oxygen, $v_{th O}\simeq 40$ km s$^{-1}$. Here we assume an intermediate value $v_{\rm ran} = 150$ km s$^{-1}$, which is close to the geometrical mean of $v_c$ and $v_{th O}$; indeed, the geometrical mean, rather than the arithmetic mean, represents the fact that in an exponential decay the initial velocity quickly decreases.
Then, $\tau_D$ is determined by fitting the observed emission profiles in the EUV.
Indeed, even the dispersion time depends on the local density as \citep{Krall} 
\be
\tau_D \approx \frac{m_T^2 v_{\rm ran}^3}{16\pi n_e e^2 q_T^2 \ln\Lambda} \, ,
\label{eq:tau_d}
\ee
so that determining $D_{xx} = \frac{1}{3}v_{\rm ran}^2\tau_D$ by fitting the observed brightness profiles, we are able to estimate $\tau_D$ and then $n_e$.

\begin{table}[htpb]
	\centering
	\begin{tabular}{l l}
	\hline
	{\bf Function or parameter} & {\bf Description} \\	
	\hline
	\hline
	$\eta(t)=\eta_0 \exp(-t/\tau_{L})$ & Ion beam ageing \\
	$\tau_{L}$ & Ion lifetime \\
	$\Phi(t') = \frac{\Phi_0}{\sqrt{2\pi}\sigma_{t_0}} \exp\left[ -\frac{(t'-t_0)^2}{2\sigma_{t_0}^2}\right]$ & Ion flux injection\\
	$\tau_\mathrm{slow}$ &  Decay time of the beam\\
	$t_0$ & Injection time \\
	$\sigma_{t_0}$ & Time interval for ion injection \\
	$x(t') = x_{00} + u_0 \tau_\mathrm{slow}\left[1-\exp(-t'/\tau_\mathrm{slow})\right]$ & Motion of the beam \\
	$u_0$ &  Beam speed along the magnetic field\\
	$D_{xx} = \frac{1}{3}v_{\rm ran}^2\tau_D$ & Diffusion coefficient\\
	$v_{\rm ran}$ & Random speed of suprathermal ions\\
	$\tau_D$ & Diffusion time \\
	\hline
	\hline
	\end{tabular}
	\caption{Description of the functions and model parameters.}
	\label{tab:tab1}
\end{table}

\section{Data analysis and comparison with modeling} \label{sec:analysis}
\subsection{Striae intensity profiles from observations}

\begin{figure}[htpb]
    \centering
    \includegraphics[width=1.0\textwidth]{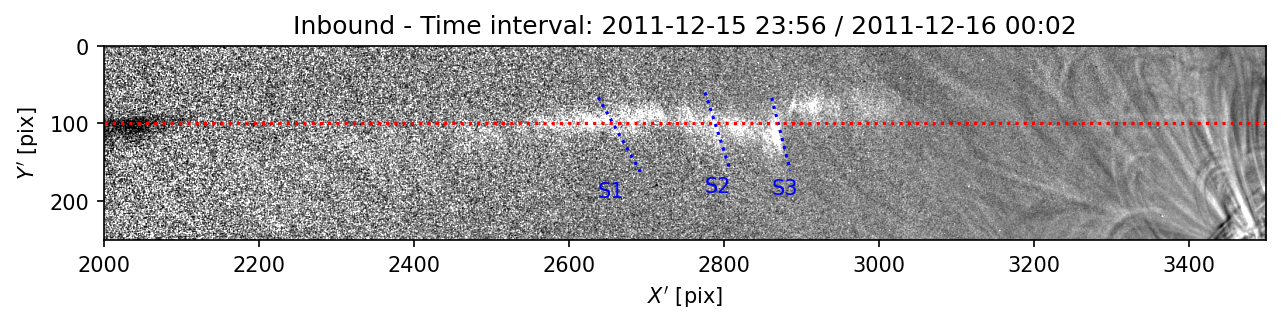}
    \caption{Re-interpolated image of the transit of Comet Lovejoy during the inbound phase as seen from SDO/AIA in 171 \AA. The image is a composition of base ratio frames taken in the time interval indicated at the top of the panel. The red dashed line marks the comet path. The analysed striae are marked by the dashed blue lines, and labelled as S1--S3. The associated animation is available (inbound.mp4). The six seconds video covers the transit of Comet Lovejoy in the corona from Dec 15, 2011, 23:46 UT to Dec 16, 2011, 00:16 UT.}
    \label{fig:remap_slits}
\end{figure}

To study the temporal evolution of some of the striae observed in 171 \AA, we re-mapped the image data-set into a new 2D grid, having the same pixel size of the original images and with the horizontal axis ($x'$) parallel to the direction of motion of the comet and the vertical one ($y'$) strictly perpendicular to its trajectory. Such a perspective is shown in Fig. \ref{fig:remap_slits}. The comet path is outlined with a dotted red line and located at $y'=100$ pix in the top panel and $y'=150$ pix in the bottom panel. Online animations are also available. 
Re-mapping is necessary to trace the striae in the corona. We noticed that the locations where the striae emission starts during the inbound phase of the comet does not lie perfectly on the dashed red line, which outlines the comet path, but it is shifted few pixels outwards. This is probably due to a systematic error in the position of the Sun's centre reported in the header of the FITS files when SDO/AIA was in off-pointing mode of observations. A physical explanation could also be found in the action of radiation pressure that pushes outwards the fragments and dust grains, before getting sublimated and ionised. However, this effect should be negligible in the present observations. In fact, such a shift is not evident in the outbound observations when SDO/AIA did return to pointing the Sun centre. 
We manually selected few bright striae, which are marked by dashed blue lines and labelled as S1,..., S6 in Fig. \ref{fig:remap_slits}. The intensity extracted from each slit and averaged over a width of 11 pixels for all the available time steps is stacked into columns forming then a time-distance (TD) map (Fig. \ref{fig:tdmap_slits}). Time is on the horizontal axis, given in minutes with respect to the time $t_c$ when the comet passes through the slit. Distance on the vertical axis, measured with respect to the position where the comet crosses the slit, is given in units of 10$^3$ km.

\begin{figure}[htpb]
    \centering
    \includegraphics[width=1.0\textwidth]{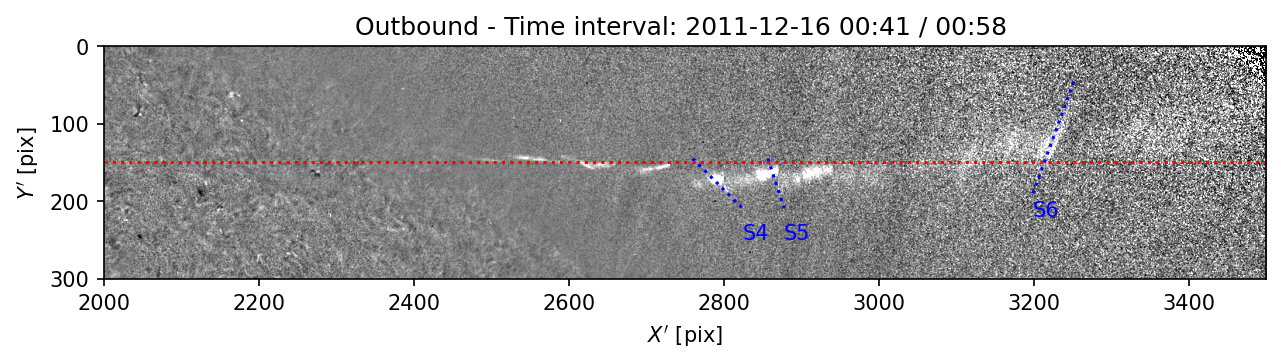}
    \caption{Same as in Fig. \ref{fig:remap_slits} but for the outbound phase of the comet trajectory. The striae marked by blue lines are labelled as S4--S6. The animation of this figure is available (outbound.mp4). The nine seconds video covers the transit of Comet Lovejoy on Dec 16, from 00:40 to 01:28.}
    \label{fig:remap_slits2}
\end{figure}

The analysis of the TD maps can provide some information in the evolution of the oxygen ion emission. The signature of a stria in a TD map is in the majority of the cases pretty diffuse and extends along the positive direction of the vertical axis. We see a sharp boundary denoting the propagating front of the ions and a bright core moving more slowly. 
In addition, the emission also slowly propagates in the negative direction of the distance axis because of dispersion effects on the moving ions. The emission in the TD maps for slits S4 and S5 is very narrow and straight, indicating that the ions are quickly transported along the magnetic field lines. The emission starts almost one minute after the comet transit and lasts for a shorter time with respect to the other striae because of the higher collision rate with coronal particles. Given a  time interval $\delta t = 2$ min and an apparent distance covered by the emission of $\Delta s \sim (20-25)\times10^3$ km, the slope of the emission in the TD maps for S4 and S5 can be associated with an average ion beam speed of 167-208 km s$^{-1}$. In the other cases, we also see some signatures coming from nearby structures, like a diffuse emission at the right side of the TD maps for the slits S2 and S6. In particular for S3, the shape of the stria signature has an inverted \lq\lq V\rq\rq-shape, which is the result of the merging of nearby striae, as it is found by inspecting the animation inbound.mp4. 

\begin{figure}[htpb]
    \centering
    \begin{tabular}{c c}
        \includegraphics[width=0.5\textwidth]{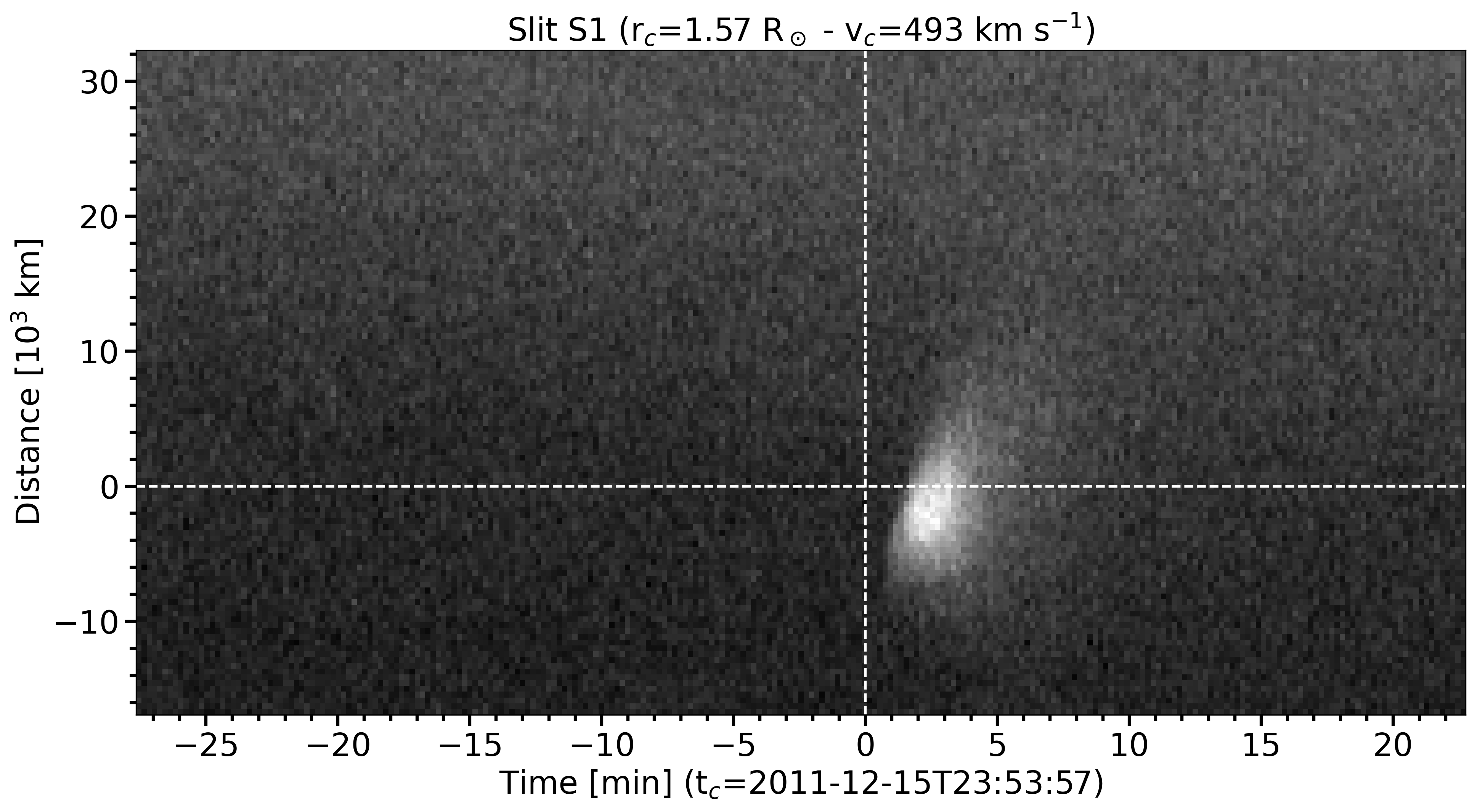}  & 
        \includegraphics[width=0.5\textwidth]{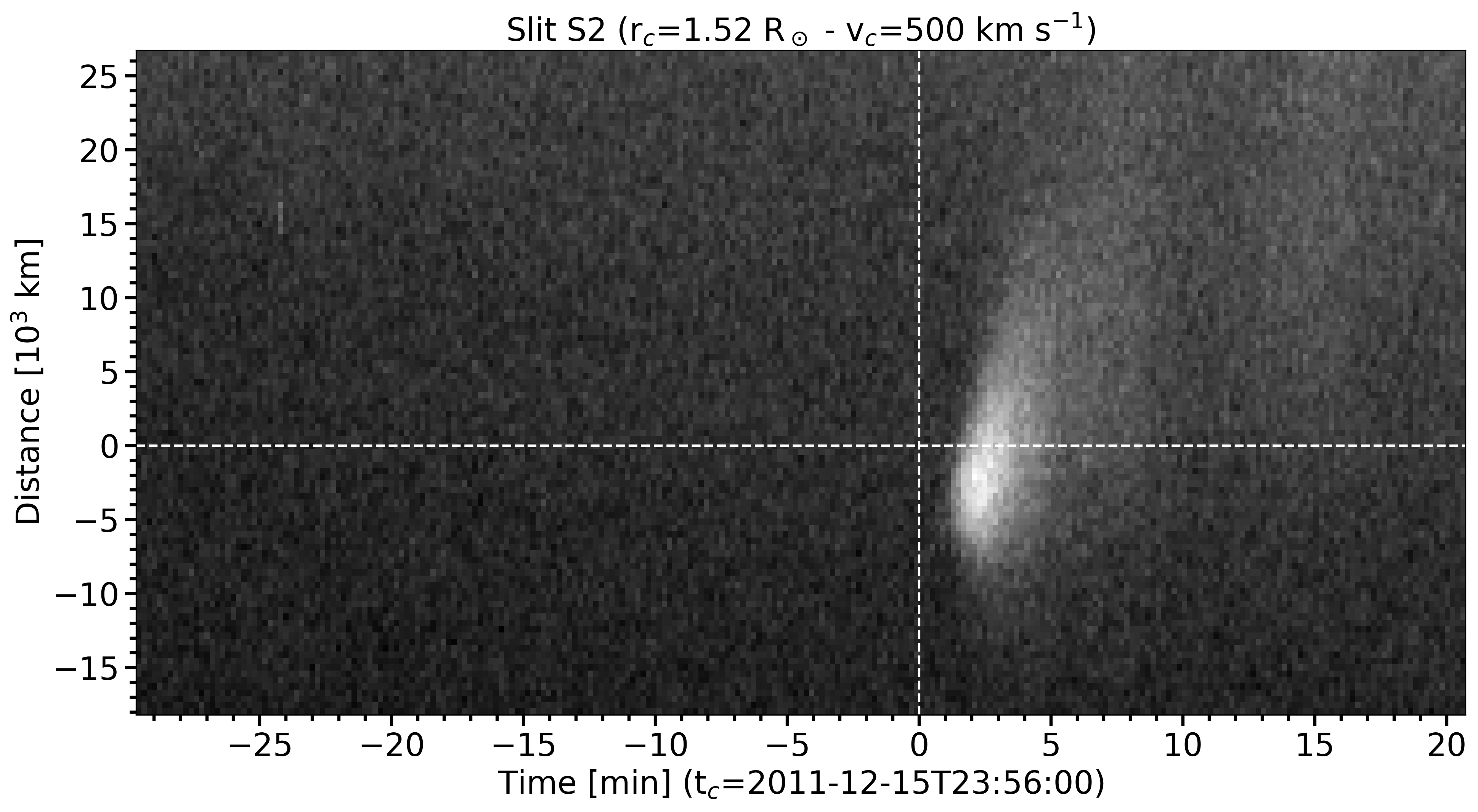}\\
        \includegraphics[width=0.5\textwidth]{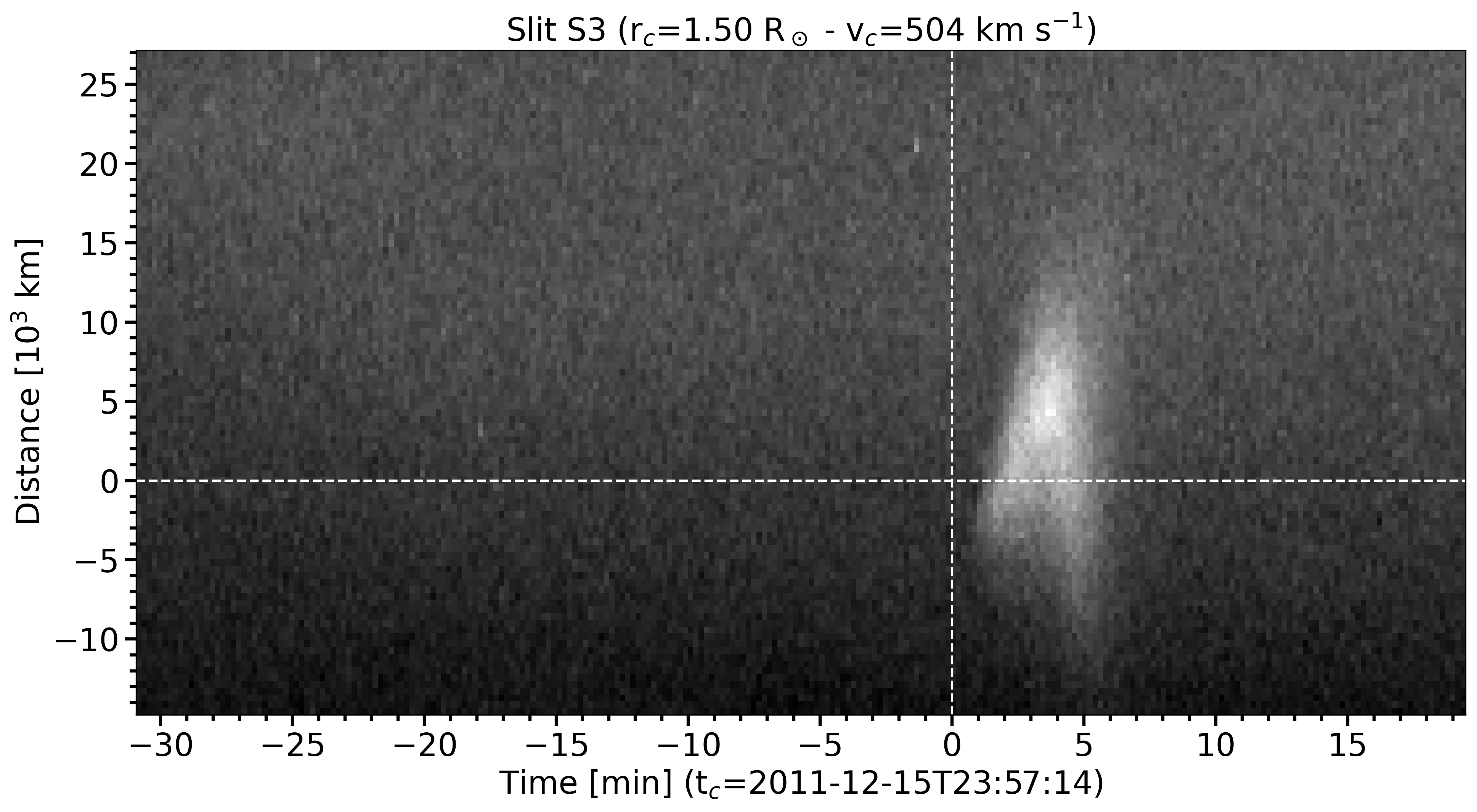}  & 
        \includegraphics[width=0.5\textwidth]{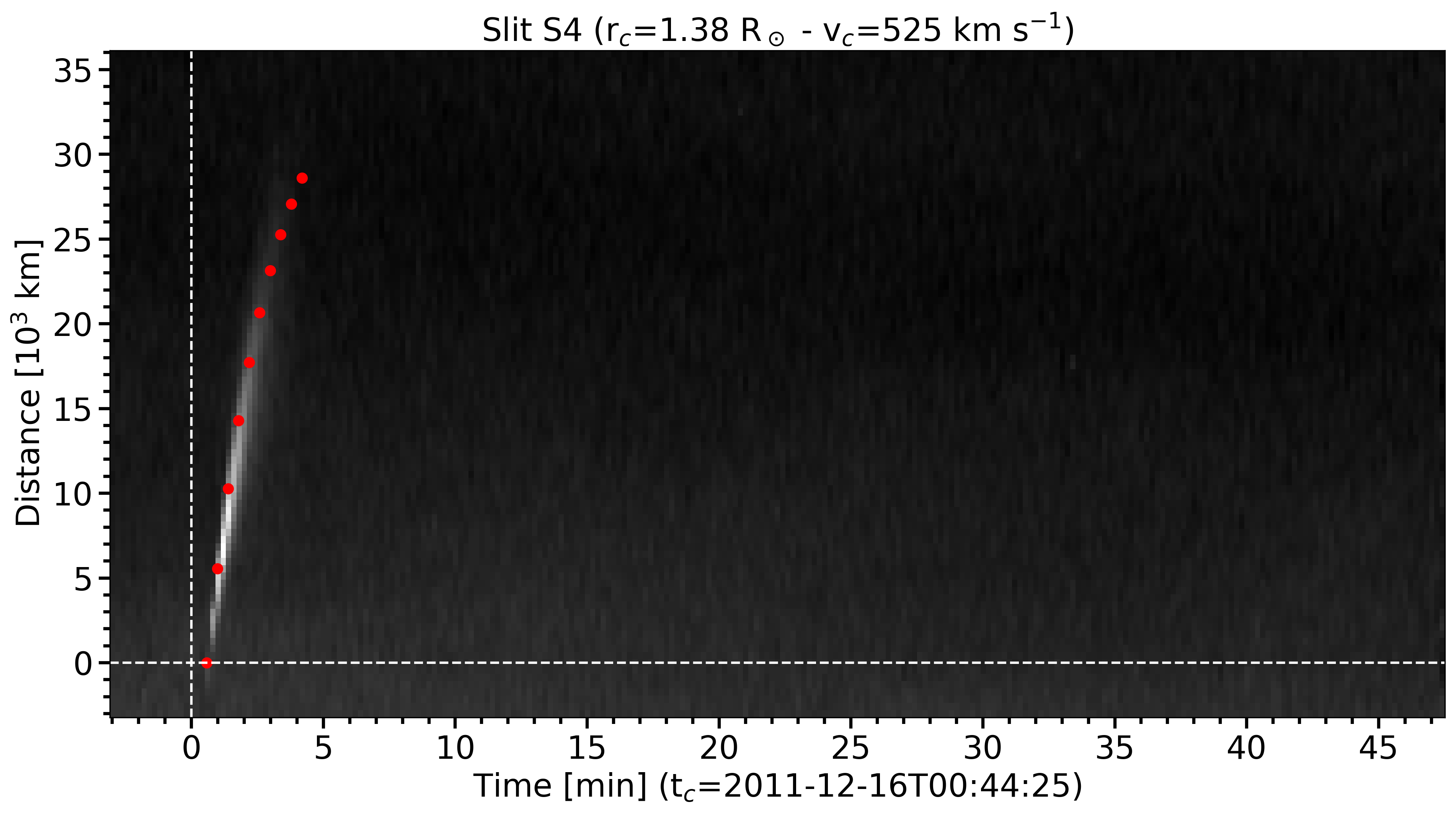}\\
        \includegraphics[width=0.5\textwidth]{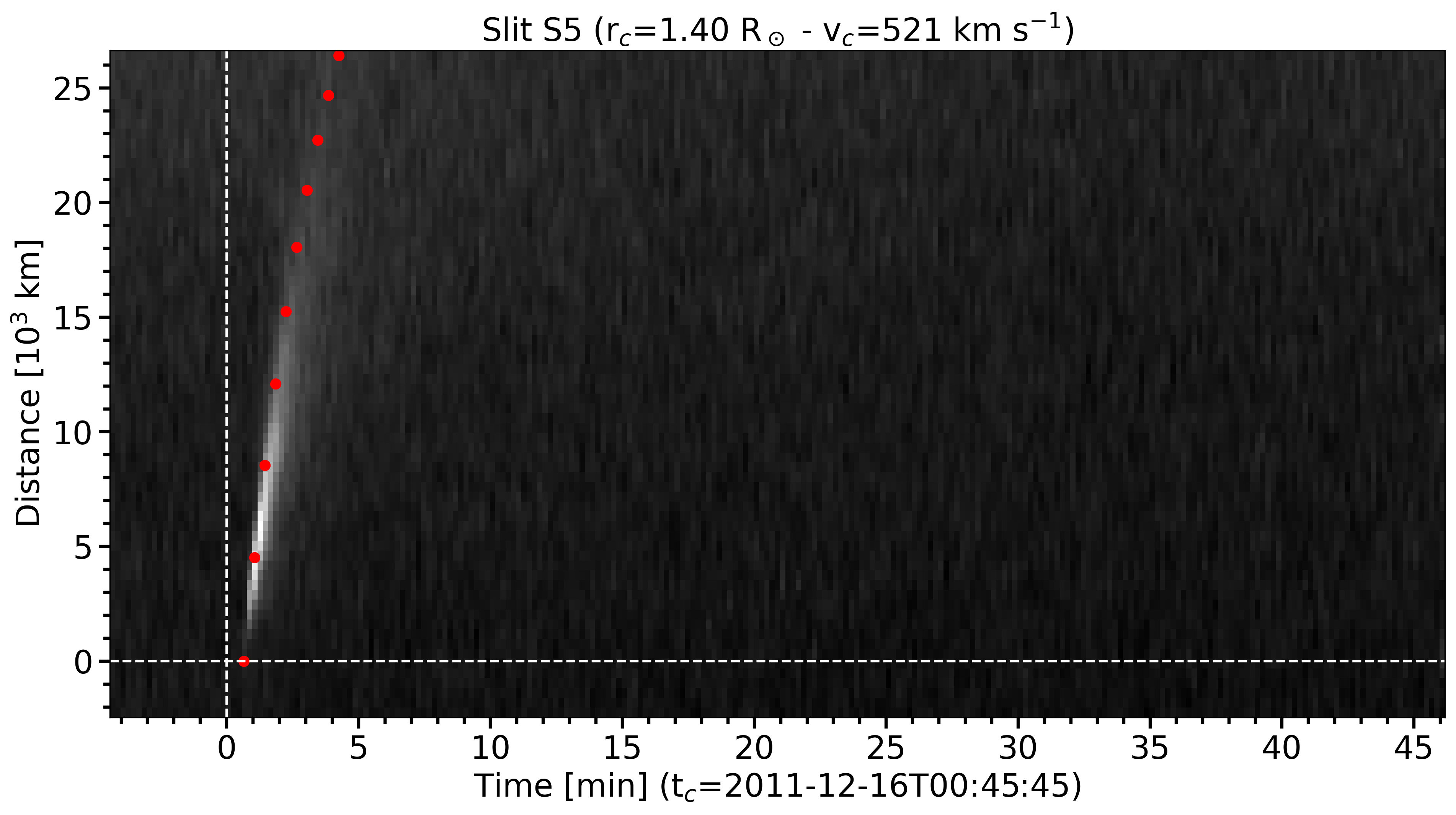}  & 
        \includegraphics[width=0.5\textwidth]{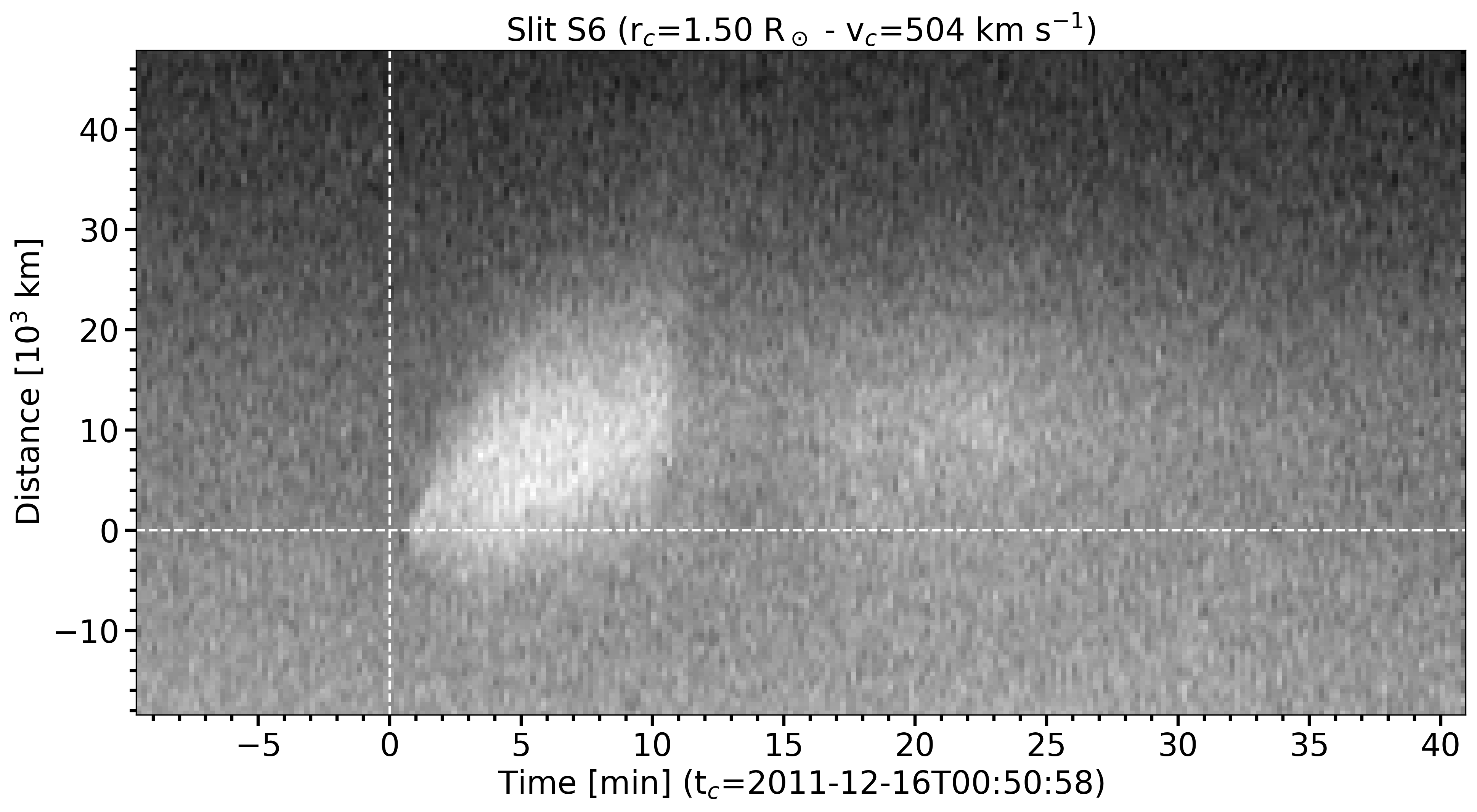}\\
    \end{tabular}
    \caption{Time distance maps relative to the slits S1--S6 marked in Fig. \ref{fig:remap_slits}. The position of the beam as a function of time obtained from Eq. \eqref{eq_beam} at time steps of 24 s are overplotted as red dots in the TD maps for slits S4 and S5. For details, see Sect. \ref{subsec:comp}.}
    \label{fig:tdmap_slits}
\end{figure}

\subsection{Comparison between observational and modeling profiles}
\label{subsec:comp}

\begin{figure}[htpb]
    \centering
    \begin{tabular}{c c}
        \includegraphics[width=0.5\textwidth]{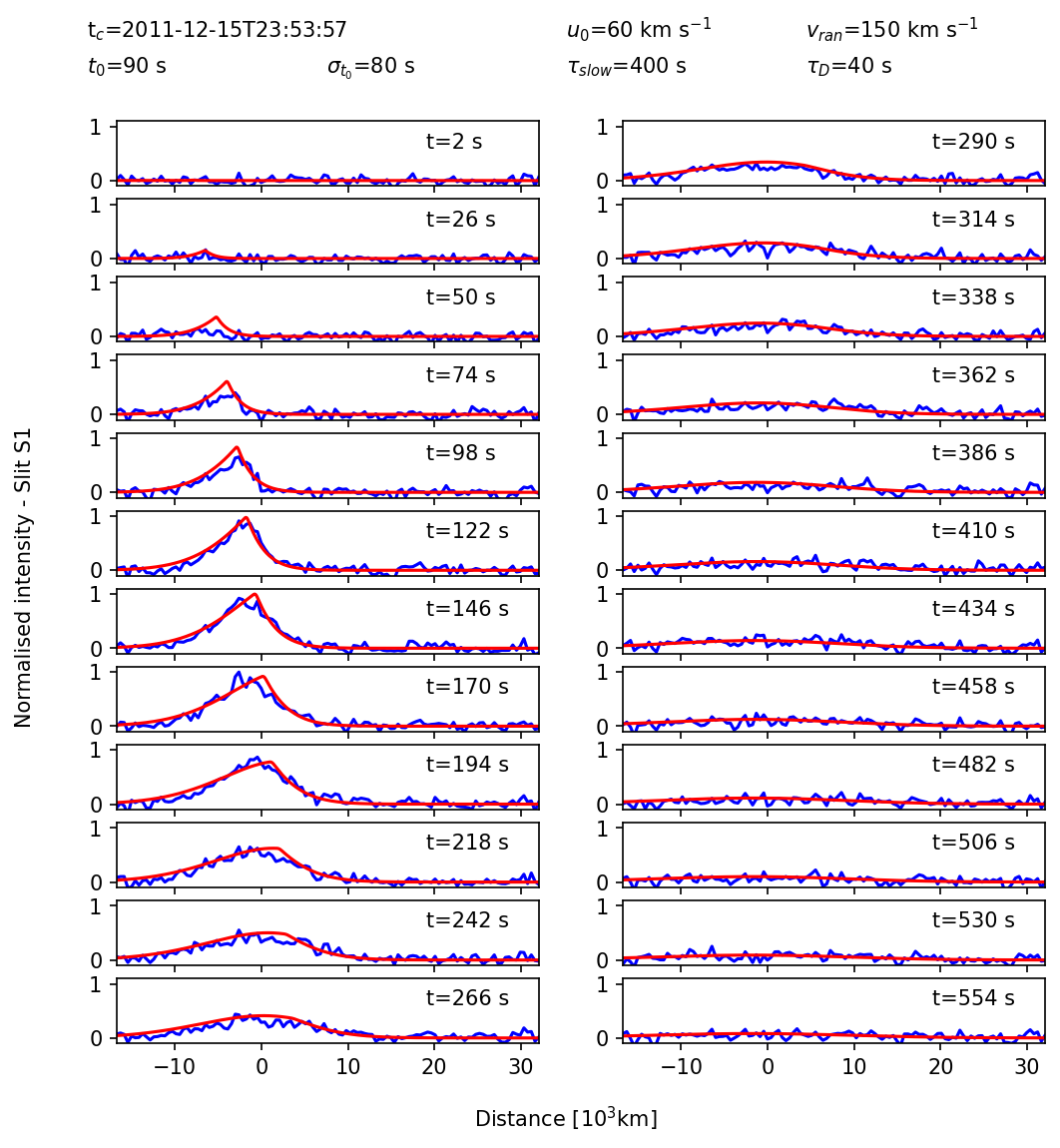} &
        \includegraphics[width=0.5\textwidth]{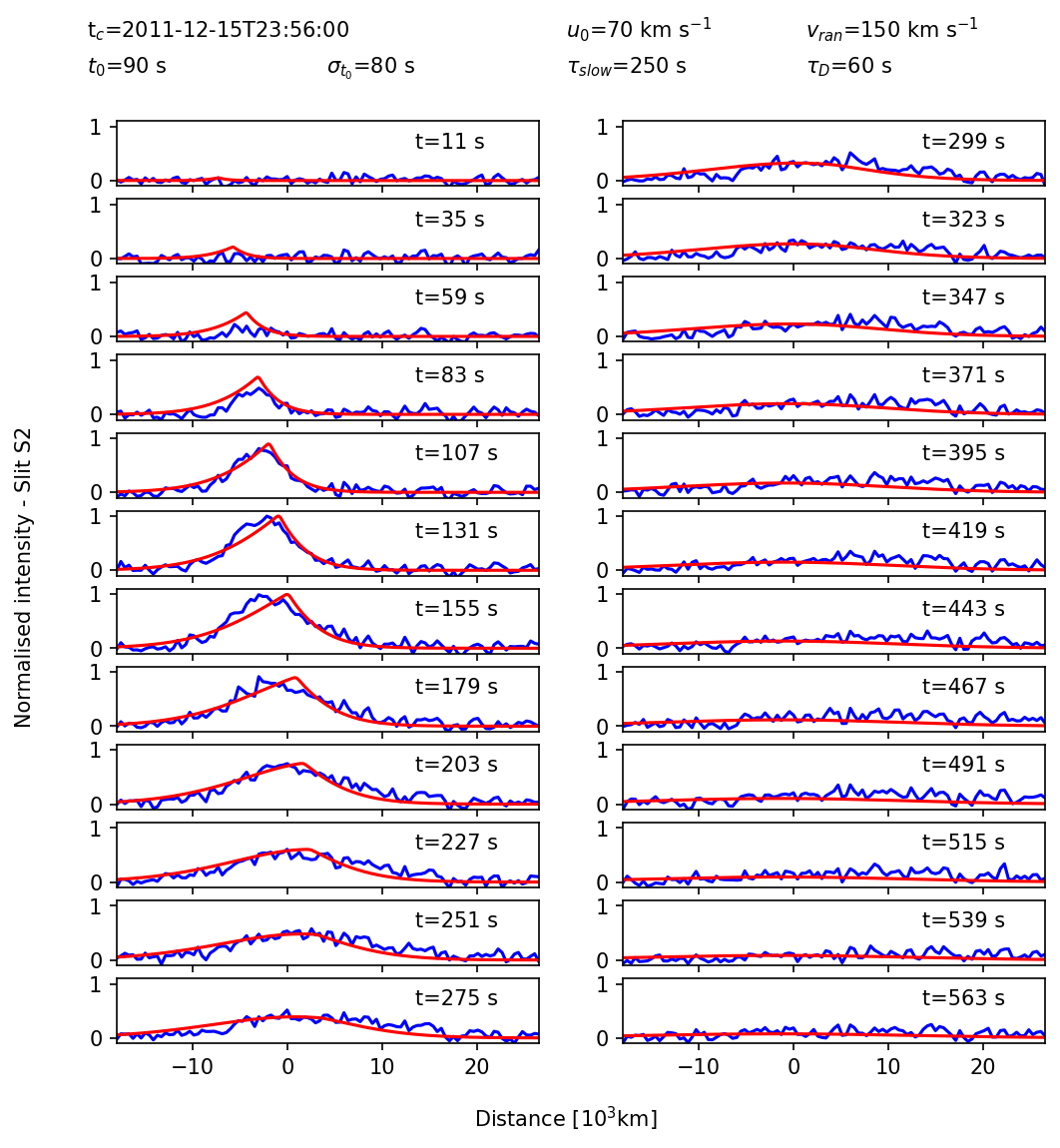} \\
        \includegraphics[width=0.5\textwidth]{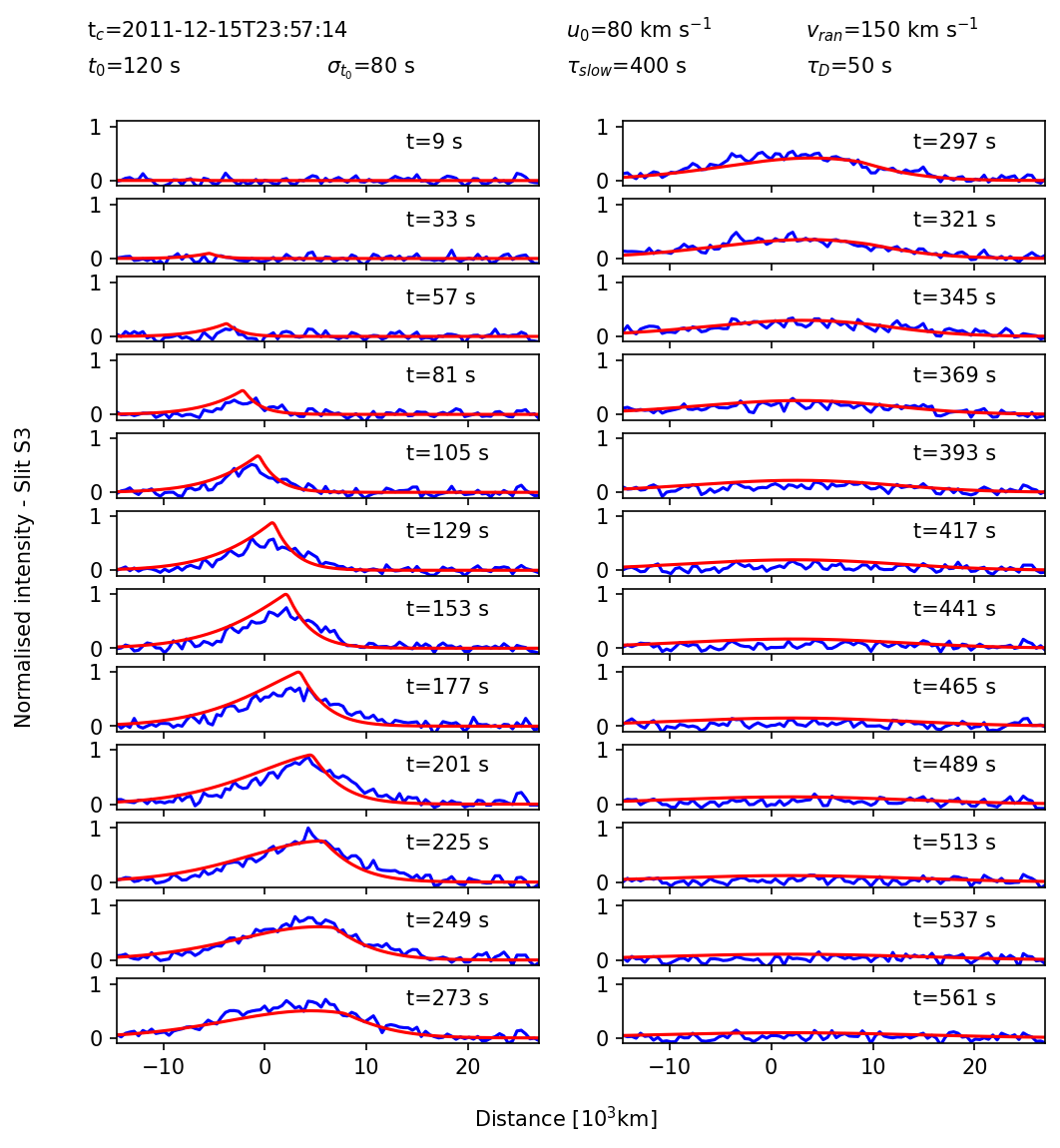} &
        \includegraphics[width=0.5\textwidth]{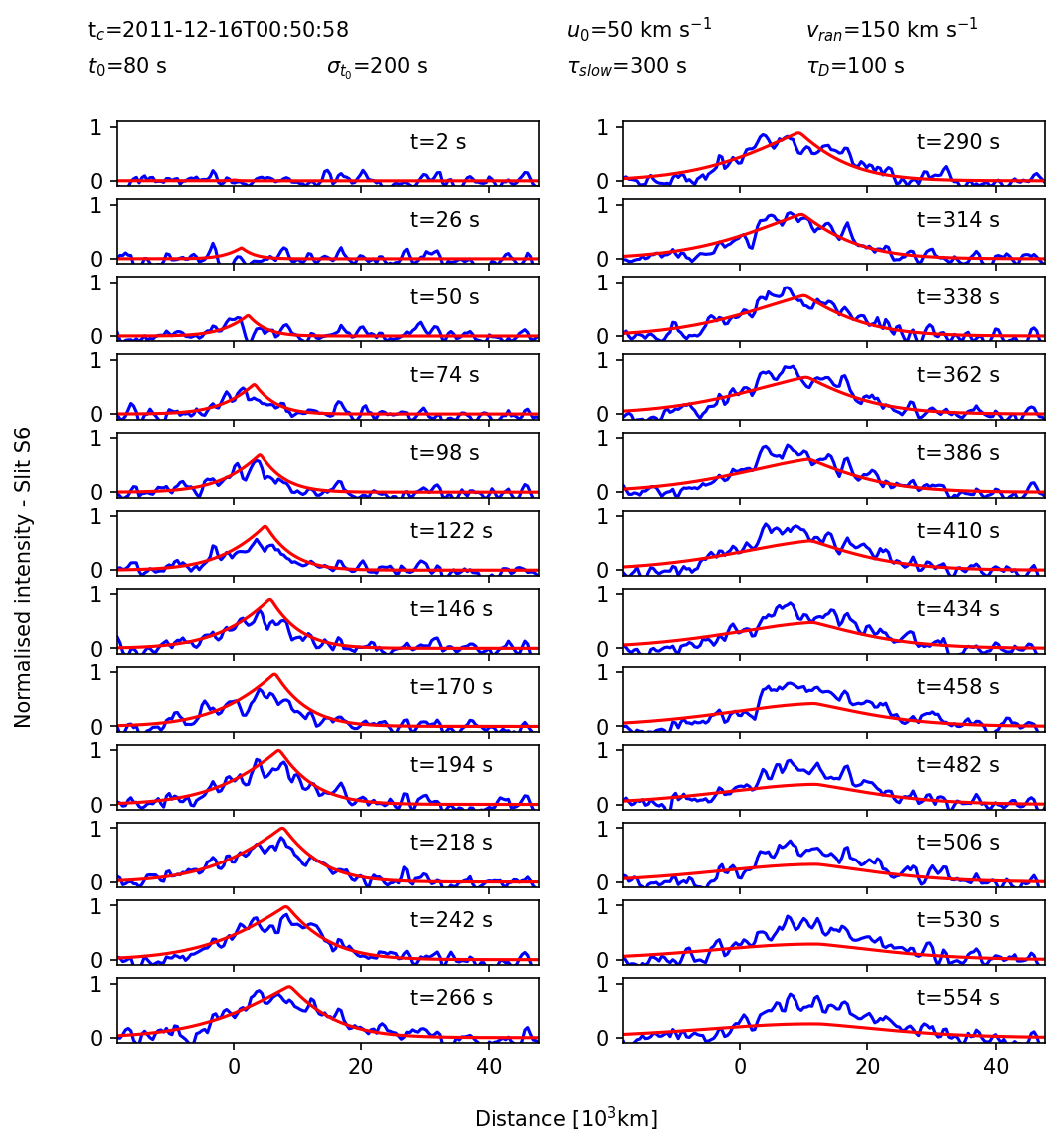} \\
    \end{tabular}
    \caption{Intensity profiles for the slits S1--S3 and S6 recorded with SDO/AIA (blue lines) and compared with the squared density profiles obtained from the modeling (red lines). }
    \label{fig:obs_mod_profiles_01}
\end{figure}

\begin{figure}[htpb]
    \centering
    \begin{tabular}{c c}
        \includegraphics[width=0.5\textwidth]{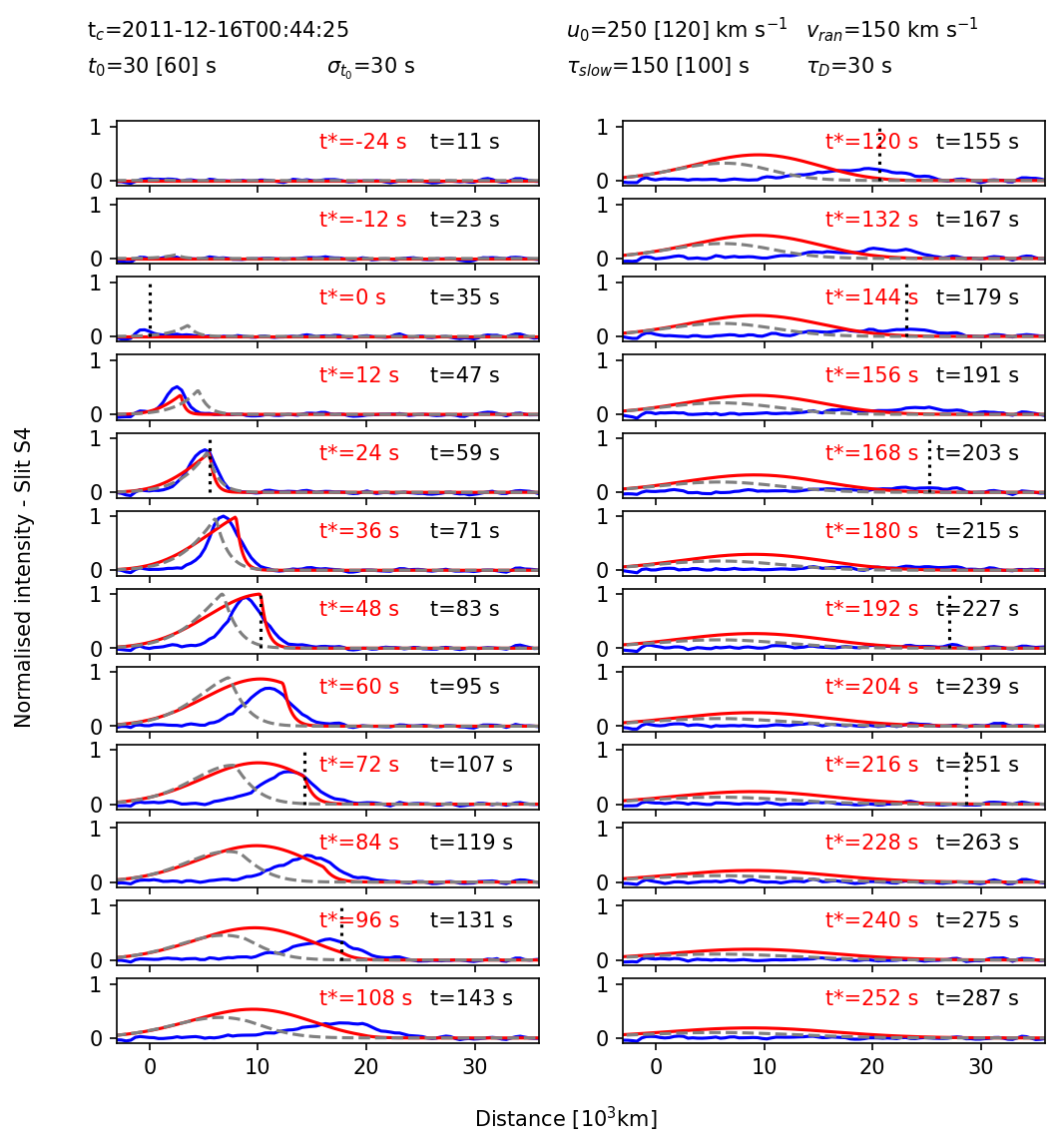} &
        \includegraphics[width=0.5\textwidth]{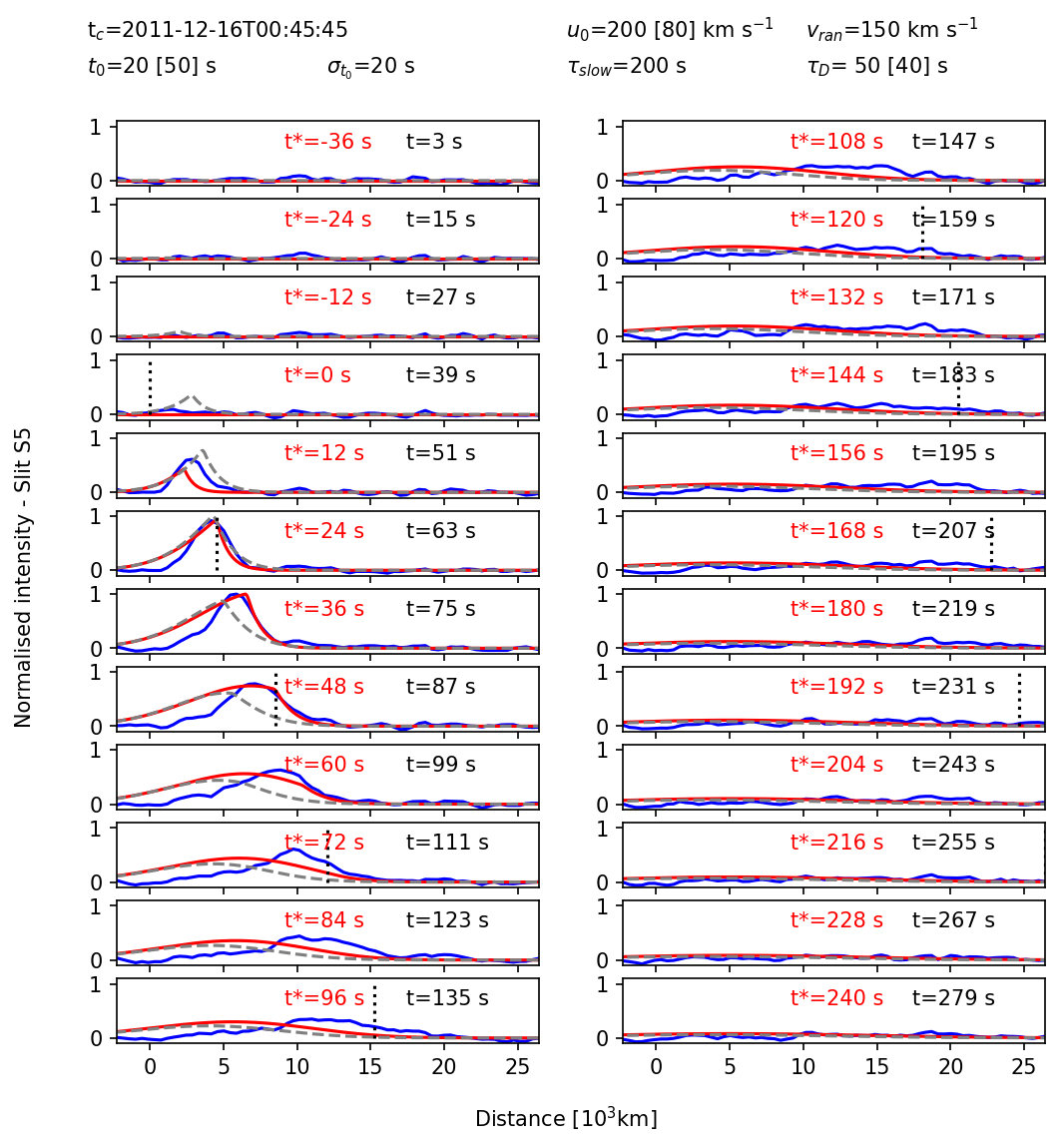} \\
    \end{tabular}
    \caption{Intensity profiles for the slits S4 and S5 as blue lines vs. squared density profiles obtained from the modeling (red line). For the density profiles we applied a temporal shift (the values of $t^\star$ in red) to better fit the rise of the intensity peak. The density profiles referred to the time of the comet transit (the values of $t$ in black) are also shown as dashed gray lines. Different values of the parameters for these profiles as given inside square brackets. The vertical dotted black line represents the position of the beam front, computed from Eq. \eqref{eq_beam} and analogous to the red points shown in the TD maps for slits S4 and S5.}
    \label{fig:obs_mod_profiles_02}
\end{figure}

By focusing on the emission signature, the intensity profile at 171 {\AA} as a function of distance at a given time (i.e., the intensity coming from a vertical line in the TD map) is compared with the squared density from Eq. \eqref{eq:eq_density_final} in Fig. \ref{fig:obs_mod_profiles_01}-\ref{fig:obs_mod_profiles_02}. The latter is integrated numerically via Simpson's rule and is computed over a spatial interval $[-5, 5]\times 10^4$ km with an integration step $\Delta x = 100$ km. The integration in time is performed in a time interval of 600 s spanning the comet transit with a time step $\Delta t =$ 12 or 24 s. 

EUV intensity and ion beam density are not directly comparable quantities. The EUV intensity from a pixel at a given wavelength $I_\lambda$ is defined as the convolution of the response function of the EUV filter $R_\lambda$ with the differential emission measure integrated along the line-of-sight $h^\prime$ \citep{2004psci.book.....A,2012SoPh..275...41B}:
\begin{equation}
    I_\lambda = \int_0^\infty R_\lambda(T) DEM(T) dT, 
\end{equation}
where the differential emission measure is defined as $DEM(T) = n^2(T, h') \frac{dh'}{dT}$ with $n$ being the free electron density of cometary origin, therefore proportional to the oxygen density \citep[Supplementary Material][]{Schrijver2012}. However, if we assume that oxygen ions are formed at a specific temperature value $T_0$ (see Table 1 of \citet{McCauley2013}, where the temperature peaks for oxygen ions based on the CHIANTI model under the hypothesis of equilibrium conditions are reported) and that the density is approximately constant through the stria column depth $h$, the DEM function can ideally be expressed in terms of a delta function $DEM(T) = n^2 \frac{dh'}{dT}\delta(T-T_0)$, hence:
\begin{equation}
    I_\lambda = \int_0^h R_\lambda(T_0) n^2(T_0) dh' = R_\lambda(T_0) n^2(T_0) \int_0^h dh' =  R_\lambda(T_0) h~n^2(T_0), 
\end{equation}
with $h$ the column depth of the stria. Therefore, provided the column depth $h$ does not vary, under such assumptions, the intensity differs from the squared density only by the multiplicative factor $R_\lambda(T_0) h$. On the other hand, the computation of the theoretical intensity fluxes (in units of photons s$^{-1}$ or DN s$^{-1}$) would require that the response function $R_\lambda$ is not based on the ordinary coronal abundances but instead on those of the mixture of coronal and cometary plasma. We avoid this issue by subtracting a background level to the stria intensity profiles and normalising these to their maximum.

The observed intensities and squared density profiles are given as blue and red lines, respectively, in Fig. \ref{fig:obs_mod_profiles_01} for slits S1--S3, S6, and in Fig. \ref{fig:obs_mod_profiles_02} for slits S4--S5. The parameters used to define the computed density profiles are educated guesses, obtained by doing several manual trials until the red line does qualitatively fit the observed profiles. More specifically, we followed these steps:
\begin{itemize}
    \item given a set of starting guessed values for the parameters, we numerically integrate Eq. \eqref{eq_density} and overplotted the computed squared density against the intensity profile;
    \item we visually checked the overlap of the profiles, both in time and space;
    \item if the profiles did not match well, we proceeded by changing the value of one parameter, we re-computed the modelled density and compared to the observed intensity profile. If the profiles did not overlap again, we re-did the procedure;
    \item when finding the best-fitting squared density profile, we changed one parameter at a time to empirically understand its effect on the trend of the profile (e.g., peak height and position, steepness of the profile, etc.).
\end{itemize}
Since the model has a large number of parameters, a Markov-Chain Monte Carlo approach would be suitable for a more robust fit \citep{2021ApJS..252...11A}. We defer this to a future work. The value of the parameters that we determined are reported in each panel of Fig. \ref{fig:obs_mod_profiles_01} and \ref{fig:obs_mod_profiles_02} and also listed in Table \ref{tab:tab2}. The intensity profiles from each slit are given with a time interval of 24 s, except for those from the slits S4 and S5 where the evolution is quicker and we preferred to show it every 12 s.   

The theoretical profiles shown in Fig. \ref{fig:obs_mod_profiles_01} are in good agreement with the observed ones for the given sets of parameters that we have determined. In the case of the intensity profiles from slits S4 and S5 (Fig. \ref{fig:obs_mod_profiles_02}), we were not able to fit well the observed intensity profiles for all the instants of time (dashed grey lines). The reasons might be different: e.g., misalignment of the slit with the stria; non-constancy of $\tau_{slow}$ because of non-negligible gravitational stratification effects; motion of the background coronal plasma due to blobs or MHD waves \citep{2021ApJ...909..202C}; the influence of plasma instabilities on the evolution of cometary plasma. Also, a varying initial speed of the ions, as the comet could be subject to some evolution  because of the solar radiation, might affect the observed intensity profiles. However, as the main reason we indicate the variable projection of the ion beam motion due to non-straight magnetic field lines. The region where S4 and S5 are located is possibly made of a system of magnetic field loops viewed mostly edge-on to the comet path and then bending and becoming more transverse with respect to the line-of-sight, as also suggested in \citet[]{Downs13} and \citet{Raymond2014}, while the striae for the other slits would be straight flux tubes. The beam motion for S4 and S5 would be inevitably affected by the curvature of the loops, and the apparent speed would be much smaller than the \lq\lq true\rq\rq one when the motion of the ion beam is almost parallel to the line-of-sight direction. For example, we notice that in the slits S4 and S5 the cloud of cometary ions did not move too much in space before getting visible. In fact, if we consider the time lag after the comet transit ($\Delta t \sim 50$ s, as reported in black in the intensity time series of Fig. \ref{fig:obs_mod_profiles_02}) and the location of the peak ($\Delta s \sim (2-4)\times10^3$ km) at its first appearance, the corresponding average beam speed would be of 40-80 km s$^{-1}$, which is in disagreement with those estimated from the slope of the bright feature in the TD maps (an acceleration of the ion beam could not be easily explained). On the other hand, we could expect that, as the ion beam descends along the loop, it progressively outlines the loop leg, which is less curved. Therefore, to mitigate the effect of the loop curvature and to avoid modelling the early phase of the ion beam propagation, when fitting the intensity profiles from S4 and S5, we considered a temporal shift and took as a reference a time instant before the appearance of the peak in the observed intensity profiles. The shifted time intervals are shown in red in Fig. \ref{fig:obs_mod_profiles_02} and labelled as $t^\star$. For these specific cases, the quantity $u_0$ will continue to be the ion beam injection speed, but no longer defined at the time of the comet transit but at the instant of time that we took as a reference. The computed and re-defined profiles are shown in red against intensity profiles (in blue) and the former profiles computed with respect to the comet time transit (dashed grey lines). We also report the value of the physical parameters for both theoretical profiles (in squared brackets those for the dashed grey lines). In general, there is an improvement in the overlap between the model and the observations: the injection beam $u_0$ is of the same order of the average speed inferred from the TD maps; the peak of the theoretical profile tends to overlap with that from the observations until $t^\star\approx 60$ s, later a mismatch in space is found probably caused by the residual curvature of the loop, whose effect was not fully removed. However, the position of the beam front, as computed by Eq. \eqref{eq_beam} and depending on $u_0$ and $\tau_{slow}$, results to be in a good agreement with the stria signature in the TD maps (Fig. \ref{fig:tdmap_slits}). Despite the above discrepancies, we provisionally consider that the values obtained for $\tau_{slow}$ and $\tau_D$ are appropriate, as a first estimate, for the striae under study.

\begin{table}[htpb]
    \centering
    \begin{tabular}{l c c c | c c c c c | c c c }
    \hline
    \hline
    Slit & $t_c$ & $r_c$ & $v_c$ & $u_0$       & $t_0$ & $\sigma_{t_0}$ &  $\tau_{slow}$ & $\tau_{D}$  & $n_{\tau_{slow}}$ & $n_{\tau_D}$ & $n_e \pm \Delta n_e$   \\
        & hh:mm:ss & [R$_\odot$] & [km s$^{-1}$]    &  [km s$^{-1}$]  &  [s]  & [s]  & [s] & [s] & [$10^7$ cm$^{-3}$] & [$10^7$ cm$^{-3}$] & [$10^7$ cm$^{-3}$]\\
    \hline
        S1              & 23:53:57 & 1.57 & 493 &  60 &  90 &  80 & 400 &  40 & 3.6 & 6.4 & $5.0 \pm 1.4$   \\
        S2              & 23:56:00 & 1.52 & 500 &  70 &  90 &  80 & 250 &  60 & 5.8 & 4.3 & $5.1 \pm 0.8$ \\
        S3              & 23:57:14 & 1.50 & 504 &  80 & 120 &  80 & 400 &  50 & 3.7 & 5.1 & $4.4 \pm 0.7$ \\
        S4$^\star$ & 00:44:25 & 1.38 & 525 & 250 &  30 &  30 & 150 &  30 &10.7 & 8.6 & $9.6 \pm 1.1$ \\
        S5$^\star$ & 00:45:45 & 1.40 & 521 &  200 &  20 &  20 & 200 &  50 & 7.9 & 5.1 & $ 6.5 \pm 1.4$ \\
        S6             & 00:50:58 & 1.50 & 504 &  50 &  80 & 200 & 300 & 100 & 4.9 & 2.6 & $3.8 \pm 1.2$ \\
    \hline
    \hline
    \end{tabular}
    \caption{Values of the parameters used to reproduce the intensity of the striae (see text for further details). The two inferred coronal electron densities, namely $n_{\tau_{slow}}$ from Eq. \eqref{eq:tau_slow}, and $n_{\tau_D}$ from Eq. \eqref{eq:tau_d}, are also shown with their combined average. The symbol $^\star$ indicates those striae whose intensity profiles are fitted in a less satisfactory way than the others, see Sect. \ref{subsec:comp}.}
    \label{tab:tab2}
\end{table}

\subsection{Determination of the electron density in the striae}

\begin{figure}[htpb]
    \centering
    \begin{tabular}{c c}
        \includegraphics[width=0.5\textwidth]{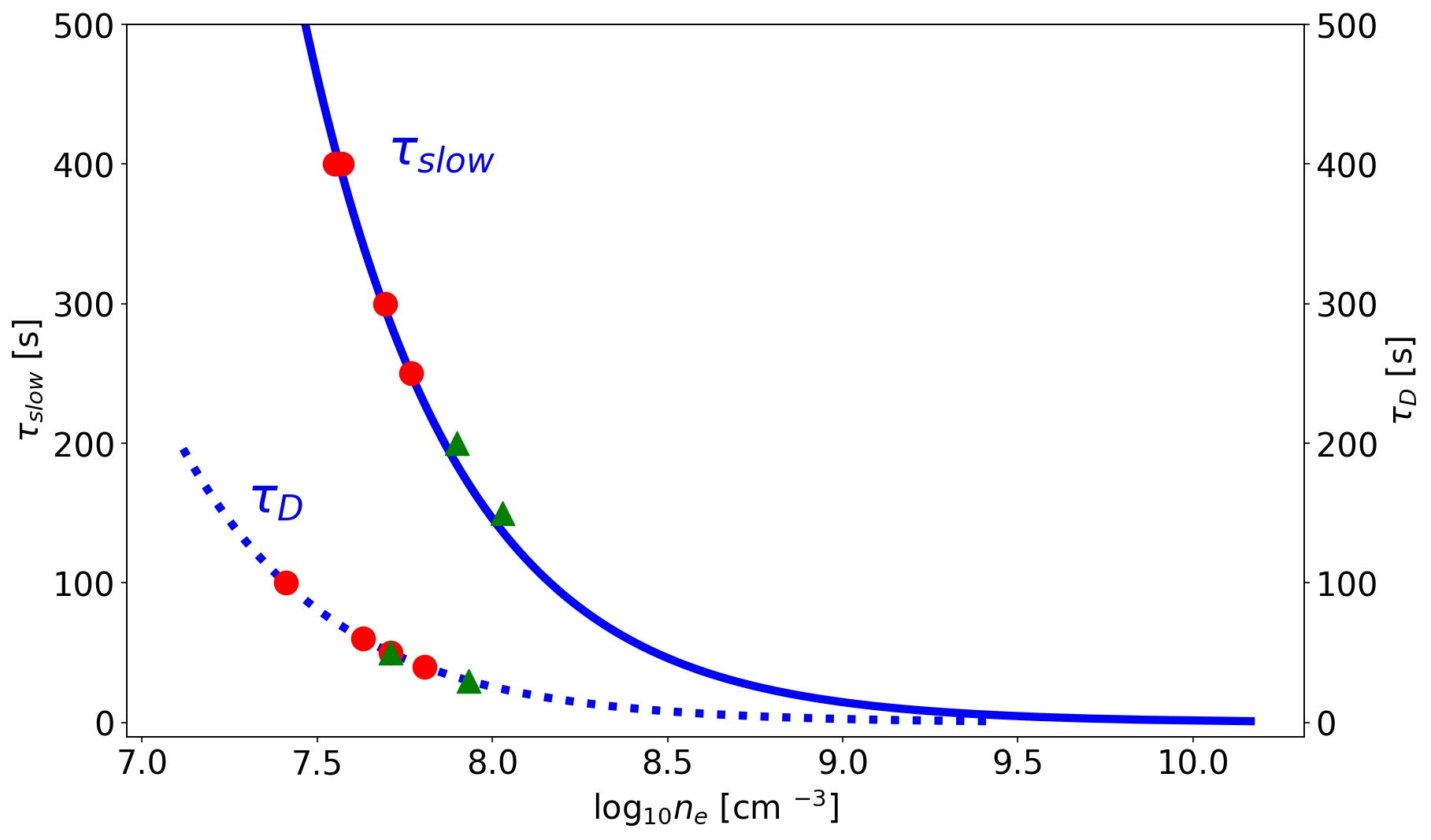} & 
        \includegraphics[width= 0.5\textwidth]{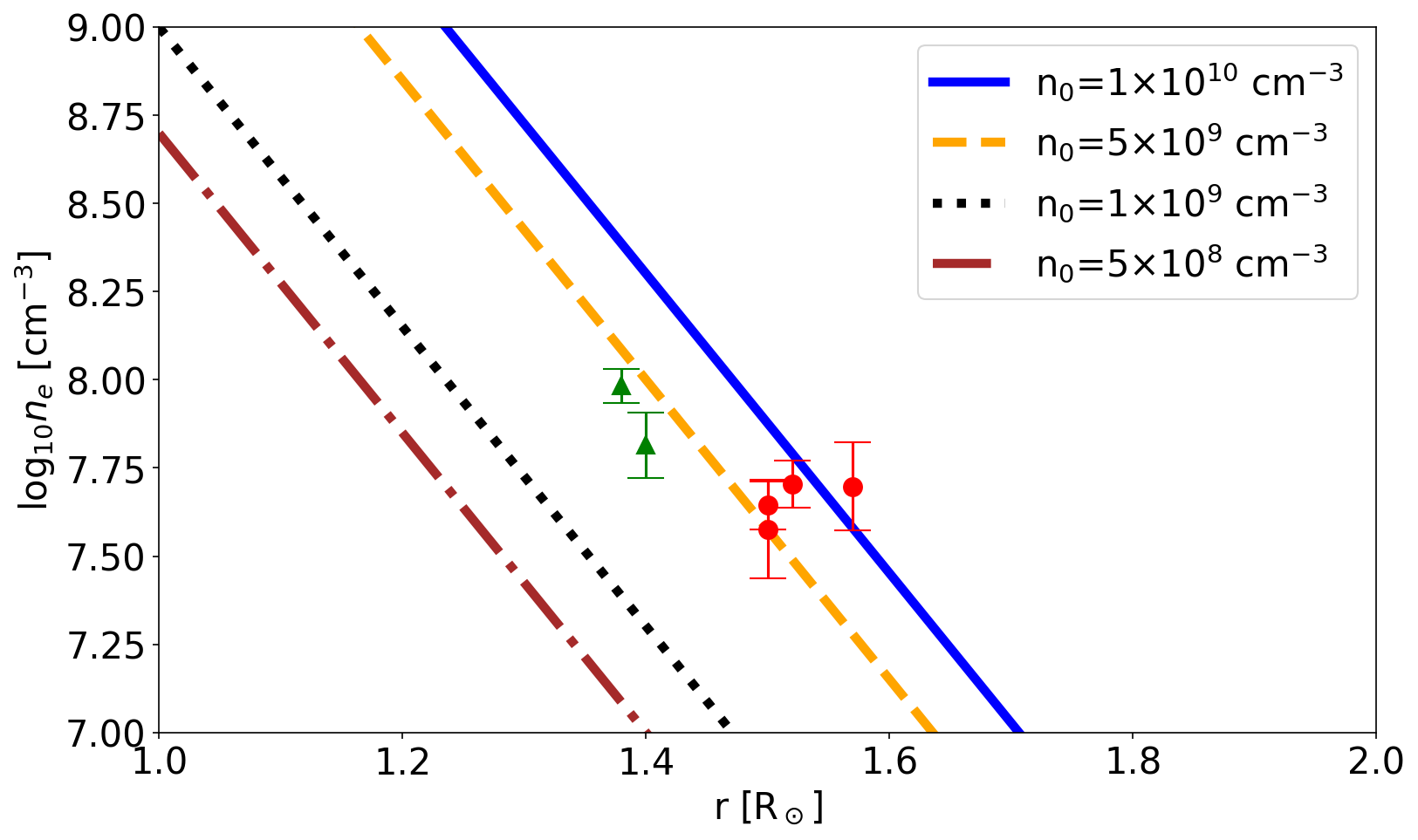}
    \end{tabular}
    \caption{Left: graph showing the trend of $\tau_{slow}$ (left axis, solid line) and $\tau_D$ (right axis, dotted line) as a function of the electron density $n_e$. The solid blue line for $\tau_{slow}$ is computed with a speed $U=$500 km s$^{-1}$ (the comet speed), while the dotted one for $\tau_D$ with $v_\mathrm{ran} = 150$ km s$^{-1}$ (the thermal speed). The two data points as green triangles refer to the slits S4 and S5. Right: electron density in logarithmic units as a function of the radial distance in $R_\odot$. The data points as green triangles refer to slits S4 and S5. The data points are compared with a hydrostatic density model for different values of the density at the solar surface $n_0$ and with $T=$1.5 MK.}
    \label{fig:tauslow_vs_ne}
\end{figure}

Once the computed profiles fit the observed ones, we used the values of $\tau_{slow}$ and $\tau_D$ to infer the electron density in the striae by using Eqs. \eqref{eq:tau_slow} and \eqref{eq:tau_d}. 
The expression of $\tau_{slow}$ depends on the injection velocity $U$ that we have assumed equal to the corresponding value $v_c$ listed in the Table \ref{tab:tab2}. The dispersion time $\tau_D$ depends on the random velocity $v_\mathrm{ran}$ that we estimated be to be 150 km s$^{-1}$ for all the analysed striae. Therefore, we provide two distinct estimates for the electron density, $n_{\tau_{\rm slow}}$ and $n_{\tau_D}$, which are also reported in Table \ref{tab:tab2}. 
In Fig. \ref{fig:tauslow_vs_ne}-left we show the dependence of $\tau_{slow}$ (solid blue line) and $\tau_D$ (dotted blue line) as a function of $n_e$. The solid line for $\tau_{slow}$ is obtained for a typical speed of the comet $U=500$ km s$^{-1}$. The data points are given as red dots for all the analysed slits, except for the slits S4 and S5, which are given as green triangles because the fit is less satisfactory for these two striae. 

The obtained estimates of density $n_{\tau_{\rm slow}}$ and $n_{\tau_D}$ are different but of the same order of magnitude. To define a single value of the coronal electron density for each  stria, we considered the average between these two estimates, i.e., $n_e = 0.5(n_{\tau_{\rm slow}} + n_{\tau_D})$ with an uncertainty simply given as the absolute semi-difference between $n_{\tau_{\rm slow}}$ and $n_{\tau_D}$  , i.e., $\Delta n_e = 0.5 | n_{\tau_{\rm slow}} - n_{\tau_D}|$. The right panel of Fig. \ref{fig:tauslow_vs_ne} shows the values of $n_e$ with the associated error bars as a function of the radial distance $r_c$ where each striae is formed. The data points are compared with a hydrostatic density profile $n(r) = n_0 \exp\left[-(r-R_\odot)/H\right]$, with  $H$ the pressure scale height defined as $H=2 k_B T_e/\mu m_p g_\odot$, with $k_B$ the Boltzmann constant, $\mu$ the mean molecular weight, and $g_\odot$ the solar gravity acceleration. We have considered $g_\odot= 274\times10^4$ cm s$^{-2}$, constant and not varying with altitude since the height of the analysed striae is $<1.5$R$_\odot$ \citep[this is the plane-parallel hydrostatic solution as presented in][]{Zucca2014}. We plotted different profiles for different values of $n_0$, the initial value of density at the solar surface. The data points tend to cluster along the lines for $n_0 = [5, 10] \times 10^{9}$ cm$^{-3}$.

\section{Discussion and conclusions} \label{sec:disc}

In this paper we present a new diagnostic method to estimate the electron density in the solar corona, at heights so low as those traveled by the sungrazing Comet C/2011 W3 (Lovejoy), which was observed in December 2011 \citep{Downs13}. In-situ measurements of the density in the solar corona at heights of hundred thousands of kilometers above the photosphere are currently not possible. In addition, measurements from remote-sensing instruments are affected by large uncertainties due to integration of emission at a certain wavelength along the line-of-sight. The method demonstrated here is based on the EUV emission of oxygen ions of cometary origin, which propagate along the local magnetic field and form the intriguing structures known as striae. Comet Lovejoy is the only comet that was observed to cross the solar corona and to survive the perihelion transit as seen from SDO/AIA \citep{Downs13} by leaving a trail of many and in some cases long-lived striae.  
The length and the spreading of striae left by Comet Lovejoy and observed in the 171 channel are related to the collision times for the slowing down of a suprathermal beam and for the diffusive spreading of suprathermal ions. These collision times directly depend on the ambient coronal density, so that a determination of them readily gives a density estimate (see Eqs. \eqref{eq:tau_slow}-\eqref{eq:tau_d}). The fitting procedure of the intensity profiles of the striae allow to determine the best values of the slowing down time $\tau_{slow}$ and the dispersion time $\tau_D$. On the other hand, a number of simplifying assumptions, like constant ambient temperature, constant ambient density, straight magnetic field lines, steady release of the ions, etc., is needed in order to extract these first results. Also, a careful data handling is necessary, since some misalignment is present between images in the sequence captured by SDO/AIA. 

The electron density was computed numerically with the use of our model and compared qualitatively with the observed intensity profile. We did several trials in order to get profiles as close as possible to the observed ones, by varying one parameter at a time. The matching between observations and model is mainly evaluated based on the position and height of the intensity peak, and the width of the tails. 
Therefore, we have shown that such an approach is useful for understanding the density structure of the solar corona with a local diagnostics once $\tau_{slow}$ and $\tau_D$ are evaluated.

In spite of some uncertainty in the obtained values of the electron density, which are in the range of $10^7-10^8$ cm$^{-3}$, we can compare them with other estimates from the literature.
\citet{McCauley2013} provides an estimate of density by considering the offset in space between the leading edge of the emission observed simultaneously in different EUV filters (which again outlines different states of ionisation for oxygen, e.g., 304 \AA~for O$^{2+}$ and 131~\AA~for O$^{5+}$) and the ionisation rate coefficients, which are linked by Eq. 2 reported in their work. By analysing the EUV emission observed around 00:46:12 UT on December 16, hence in a region of the corona included between our slits S4 and S5, they found a density of $\sim1.4 \times 10^8$, which is \sout{very} close to our values for these slits.
\citet{Raymond2014} have also given some estimates of the local electron density for the same region of interest with two different approaches: the former based on the ratio between the ionisation time of the cometary hydrogen and oxygen atoms and the typical crossing time of a stria by the comet, the latter by taking into account the emission measure, the filling factor and a typical coronal length scale observed in the striae. They give an upper limit value for the inter-striae regions of $2.7\times10^6$ cm$^{-3}$, while within a stria the lower and upper limit are evaluated as $1.7\times10^7 - 1.4\times10^8$ cm$^{-3}$, in reasonable agreement with our results.
On the other hand, our values of $n_e$ are somewhat larger than those found at the same coronal heights by \citet{Zucca2014} for both active regions and the quiet sun (see their Figure 10). Those density estimates were obtained by interpolating the densities obtained at lower altitudes by SDO/AIA and at higher altitudes by SoHO/LASCO by means of the emission measure, which is integrated along the line of sight. Therefore, those coronal densities are average values while our estimates are local values. The fact that the local values are somewhat larger than the average values agrees with the idea that coronal density is strongly spatially structured \citep[for the physical causes of such a structuring see the discussion in][]{Raymond2014}, and that the observation of striae is due to the fact that the comet is crossing regions of quickly changing density, with denser regions giving rise to larger erosion of cometary material and therefore brighter UV emission. Such a density fine structuring could also be attributed to the bimodal structure of the lower solar wind detected higher up from 6 to 21 solar radii \citep{2018PhRvL.121g5101C}.

The determination of the coronal electron density that we propose is complementary to differential emission measure analysis techniques that are widely used \citep{2012A&A...539A.146H,2015ApJ...807..143C} but not applicable to the case of cometary plasma because the condition of ionisation equilibrium is not satisfied. Moreover, it is crucial in the context of coronal seismology \citep{2020ARA&A..58..441N}. In fact, the striae left by the comet were not static features but exhibited an apparent transverse motion, perhaps an evidence of propagating kink waves \citep{2021SSRv..217...73N} in the flux tubes outlined by the striae. The transverse motion of the striae is evident in the region that encompasses the slits S2 and S3 during the inbound transit of the comet, and the region around slit S6 during the outbound phase (see the animation anim.m4v of Fig. \ref{fig:remap_slits}). Such a prospect opens a further possibility for the inference of the coronal magnetic field in the low corona by MHD seismology. 

The inhomogeneity of the plasma density revealed by the cometary striae could be due to the different densities in different magnetic flux tubes linked to solar granulation cells and small scale reconnection, which may inject different amounts of plasma in adjacent flux tubes through spicules or jets \citep{2009SoPh..259...87N, 2010AnGeo..28..687N}. These density variations can also be related to the density fine structures seen in solar eclipses images acquired from the ground, in particular in the polar regions \citep[e.g., see ][]{2009ApJ...702.1297P,2014ApJ...785...14D}. It is notable that the density structures seen in white light images have low contrast because of the line-of-sight integration, but the brightness contrast along the comet Lovejoy path reported by \citet{Raymond2014} suggests that the density ratio between adjacent flux tubes is of order six or more. Our study is consistent with those density ratios, and this gives interesting clues to the understanding of pressure balance in the low beta corona and to the dispersion of MHD waves propagating through such inhomogeneities \citep{2013A&A...549A..54M,2014A&A...569A..12N,2020ARA&A..58..441N}. Furthermore, the density fine structuring of the solar corona could also be studied with the upcoming observations of Solar Orbiter/Metis in white light and in the Lyman-$\alpha$ line \citep{2020A&A...642A..10A}, with the Association of Spacecraft for Polarimetric and Imaging Investigation of the Corona of the Sun (ASPIICS) on board the Project for On-Board Autonomy (PROBA-3) \citep{2021A&A...652A...4S}, and possibly be related to in-situ measurements with Parker Solar Probe \citep{2019Natur.576..228K,2021ApJ...920L..14T}. 

The model that we presented here explores the role of collisions in the diffusion of cometary oxygen ions and in the slowing down of the ion beam. As already mentioned, other physical factors might be at work in the evolution of cometary ions along the coronal flux tubes. For example, the diffusion along the field could also be affected by turbulence (e.g., whistler turbulence or Langmuir turbulence). Furthermore, in this work we did not investigate the influence of the coronal density on the oxygen lifetime $\tau_L$ (where $\tau_L=(n_e q_i)^{-1}$, with $q_i$ the ionisation rate), which was kept as a fixed parameter in our analysis. It would be also interesting to apply this analysis to striae observed in the other EUV channels of SDO/AIA, in order to better constrain the density values of the solar corona. We will study all these aspects in follow-up works. 
\\
\\
\\

Data are courtesy of the SDO/AIA team. G.N. acknowledges support from the \lq\lq Rita Levi Montalcini 2017\rq\rq fellowship contract funded by the Italian Ministry of University and Research for the project {\it \lq\lq Probing the heliospheric and near-Sun environments with comets\rq\rq}. V.M.N. acknowledges support from the STFC Consolidated Grant ST/T000252/1. Work of MD was supported by the Grant Agency of the Brno University of Technology (grant FSI-S14-2290). We thank the referee for his suggestions, comments and careful reading of our work.

\bibliography{references}{}
\bibliographystyle{aasjournal}

\end{document}